\documentclass[journal,onecolumn,Manuscript]{IEEEtran}

\ifCLASSINFOpdf
  
\else
  
\fi

\usepackage{amsmath}
\usepackage{amsthm}
\usepackage{cases}
\usepackage{amsfonts}
\usepackage{cite}
\usepackage{amssymb}
\usepackage{enumerate}
\usepackage{graphicx}
\usepackage{float}
\usepackage{caption}
\usepackage{verbatim}
\usepackage{dblfloatfix}
\usepackage[cmintegrals]{newtxmath}
\usepackage[ruled,linesnumbered]{algorithm2e}
\usepackage{color}

\newtheorem{theorem}{Theorem}
\newtheorem{lemma}{Lemma}
\newtheorem{definition}{Definition}
\newtheorem{corollary}{Corollary}
\newtheorem{remark}{Remark}

\begin{document}

\title{Integrated Sensing and Communication with Distributed Rate-Limited Helpers}

\author{
  \IEEEauthorblockN{Yiqi Chen\IEEEauthorrefmark{1},
  Holger Boche\IEEEauthorrefmark{1},
  Tobias J. Oechtering\IEEEauthorrefmark{2},
  Mikael Skoglund\IEEEauthorrefmark{2}}\\
  \IEEEauthorblockA{\IEEEauthorrefmark{1}%
Technical University of Munich,  
80290 Munich, Germany, 
\{yiqi.chen,boche\}@tum.de}\\
\IEEEauthorblockA{\IEEEauthorrefmark{2}%
KTH Royal Institute of Technology,  
100 44 Stockholm,
Sweden, 
\{oech,skoglund\}@kth.se}
}

\markboth{Journal of \LaTeX\ Class Files,~Vol.~14, No.~8, August~2015}%
{Shell \MakeLowercase{\textit{et al.}}: Bare Demo of IEEEtran.cls for IEEE Journals}

\maketitle
\thispagestyle{empty}
\pagestyle{empty}
\begin{abstract}
This paper studies integrated sensing and communication (ISAC) systems with two rate-limited helpers who observe the channel state sequence and the feedback sequence, respectively. Depending on the timing of compressing and using the state information, our proposed coding scheme gives an inner bound of the capacity-compression-distortion tradeoff region. The tradeoff is realized by sending part of the state information at the beginning of the transmission to facilitate the communication and compressing the remaining part together with the feedback signal. The inner bound becomes tight bounds in several special cases.
\end{abstract}


%
\IEEEpeerreviewmaketitle

\section{introduction}
It is well known that for a state-dependent discrete memoryless channel (SD-DMC), knowing states at the encoder and decoder causally/noncausally helps the message communication\cite{shannon1958channels}\cite{gel1980coding}, and the feedback link does not improve the channel capacity. 

However, things are different when state information transmission is also part of the communication system, in which the encoder observes the channel states and tries to communicate the state information to the decoder. This model is known as the state amplification problem\cite{kim2008state}. It turns out that for the joint state and message communication, a feedback link improves the message communication capacity when the state construction satisfies some required fidelity\cite{bross2017rate}. More results about the state amplification can be found in \cite{sutivong2005channel,kim2008state,choudhuri2013causal,chia2013estimation}. In addition, \cite{zhang2011joint} studied the joint message communication and state estimation when the encoder does not have access to state information.

The emerging applications in wireless communication system such as indoor localization, radar sensing for vehicle network, autonomous driving and robotics\cite{liu2022survey,schwenteck20236g} pose new requirements and challenges for the next generation wireless communication technologies. As a promising feature of the 6G network, integrated sensing and communication designs the communication and sensing systems jointly. The two systems share the same frequency band and hardware to improve spectrum efficiency and reduce the energy cost of the hardware\cite{liu2022survey}. The ISAC model moves the state estimation task from the decoder side to the encoder side, and the first fundamental information-theoretic result on the communication rate-state estimation distortion tradeoff was provided in \cite{kobayashi2018joint}. The model was then extended to multi-user case\cite{ahmadipour2022information,ahmadipour2023information}, continuous alphabet cases\cite{xiong2023fundamental}, ISAC with security and privacy constraints\cite{gunlu2023secure,wang2024covert,chen2024distribution} and error-exponent analysis\cite{joudeh2022joint,chang2023rate}.

Despite significant efforts dedicated to ISAC, its full potential remains underexplored, particularly within communication networks. Advances in wireless communication technologies have enabled each node in a network to function not only as a transmitter or receiver but also as a sensor. While remote sensing of a target poses challenges for estimators, nodes surrounding the target can act as sensors, observing or even predicting the target's state with high accuracy. This information can be leveraged to enhance sensing quality when these sensor nodes utilize their communication capabilities. A practical example of this scenario is an autonomous vehicle network, where vehicles, infrastructure, and base stations collectively serve as nodes. For instance, when a base station attempts to estimate the location of a target vehicle, nearby vehicles and infrastructure can observe and predict the target's movement in subsequent frames, encoding this information to the base station. Furthermore, feedback mechanisms should extend beyond echo signals. Protocols such as Automatic Repeat Request (ARQ) provide encoded feedback to the sender, thereby improving communication quality.

To model the aforementioned scenario, we consider a more general model such that the estimator is assisted by some rate-limited helpers. SD-DMCs with a rate-limited helper that observes the state information were investigated in \cite{lapidoth2023state} for causal state observation and \cite{rosenzweig2005channels} for noncausal state observation. Here, it should be emphasized that the definition of 'rate-limited helper' in this paper is different from that in \cite{lapidoth2023state}. In \cite{lapidoth2023state}, the helper is allowed to send a symbol in each time frame with alphabet $\mathcal{T}$ such that $2\leq |\mathcal{T}| < |\mathcal{S}|$, and the limited rate of the helper is defined by
\begin{align}
    \log |\mathcal{T}|,
\end{align}
while the rate limitation in this work is from a lossy compression perspective, which is the same as \cite{rosenzweig2005channels}. The line of work on SD-DMCs with helpers also includes communication with a message-cognizant helper who knows the message to be transmitted in advance. It hence is able to cooperate with the sender to achieve a higher rate\cite{lapidoth2012multiple2} and communication with a cribbing-helper who observes the channel input in each time frame\cite{lapidoth2024state}.

The rate-limited helpers in this paper incorporate two rate-limited encoders that observe the noncausal state sequence and the feedback sequence, respectively. It is obvious that the feedback signal is helpful for the estimation, and the state information is useful for both the estimation and communication. We consider the capacity-compression-distortion tradeoff for this model. The compression is part of the tradeoff due to the different timing that the sender uses the state information. Under a given level of the distortion constraint, since the state encoder observes the state noncausally, it can send a lossy description of the state sequence to the sender at the beginning of the transmission block to improve the communication rate between the sender and the receiver. On the other hand, the state and feedback sequences are a pair of correlated sequences, and the helpers can achieve a lower compression rate by compressing them together. This happens at the end of the transmission block due to the strict causality of the feedback signal, and hence cannot help the message communication. In this case, the state helper encodes the state information as if it observes the state causally. The proposed coding scheme in this paper is a tradeoff between the noncausal and causal state helpers. It sends part of the information to facilitate the communication and compresses the remaining part together with the feedback signal. 

The main result of this paper is an inner bound of the considered model achieved by using the strategy we mentioned in the previous paragraph.  Although the tightness of the general inner bound is still an open problem, we consider a special case in which the channel state information is known at the receiver side, and the feedback signal is required to be reconstructed at the encoder side losslessly. We provide inner and outer bounds when the feedback encoder is rate-unlimited and rate-limited. The tightness of these results heavily depends on the decoder's accessibility to the two helpers. For a rate-unlimited feedback encoder, the inner and outer bounds meet each other when the state encoder is message-cognizant. Here, we adopt a different definition of a message-cognizant encoder. The state encoder is informed of the decoding result by the decoder at the end of the transmission block instead of knowing it at the beginning of the transmission as in \cite{lapidoth2023state,lapidoth2023state2}. The analysis of the rate-limited feedback encoder is more complex. Since the state encoder always sends a description to the message sender to facilitate the communication, this description is used as common side information at the state encoder for further compression, but can only be decoder side information for the feedback encoder. This fact results in two different inner bounds when the joint distribution of the state and channel output is decomposable. It is known that a pair of random variables with a decomposable joint distribution have a common component\cite{witsenhausen1975sequences}. When the joint distribution of the state and channel output is decomposable, the compression at the two helpers can be regarded as a distributed compression for sources with common components\cite{wagner2011distributed}, and the two achievable regions depend on who sends the common description. These two regions can be unified when the receiver has full access to the feedback encoder since, in this case, the source observed by the state encoder is a deterministic function of the feedback encoder (receiver). Similar to the rate-unlimited case, the inner and outer bounds meet each other when the state encoder is message-cognizant.

The results of this paper establish fundamental performance bounds for the considered model. While these results do not constitute proof-of-concept experiments, they serve as critical benchmarks for assessing the efficacy of algorithms proposed for this model and their corresponding simulation outcomes. These bounds provide a theoretical foundation for evaluating the quality and practical relevance of algorithmic contributions in this domain.
 The rest of the paper is organized as follows. Section \ref{sec: models and results} provides the model definitions and main results. Section \ref{sec: examples and applications} gives some numerical examples and applications of the main results. The direct and converse parts of the proof of our main results are provided in Sections \ref{sec: proof of the general inner bound} and \ref{sec: converse proofs}, respectively.

\section{Models and Results}\label{sec: models and results}
Throughout this paper, random variables, sample values, and their alphabets are denoted by capital, lowercase letters, and calligraphic letters, respectively, e.g. $X$, $x$, and $\mathcal{X}$. Symbols $X^n$ and $x^n$ represent random sequence and its sample value with length $n$. The distribution of a random variable $X$ is denoted by $P_X$, and the joint distribution of a pair of random variables $(X,Y)$ is denoted by $P_{XY}$. The expectation of a function of the random variable $X$ is written by $\mathbb{E}_X\left[ f(X) \right]$. The set of integers from $1$ to $N$ is denoted by $[1:N]$. The indicator function of an event $A$ is defined by $\mathbb{I}\{A\}$ such that $\mathbb{I}\{A\}=1$ if $A$ is true and $0$ otherwise.
\subsection{Models}
\begin{figure}[h]
    \centering
    \includegraphics[scale=0.5]{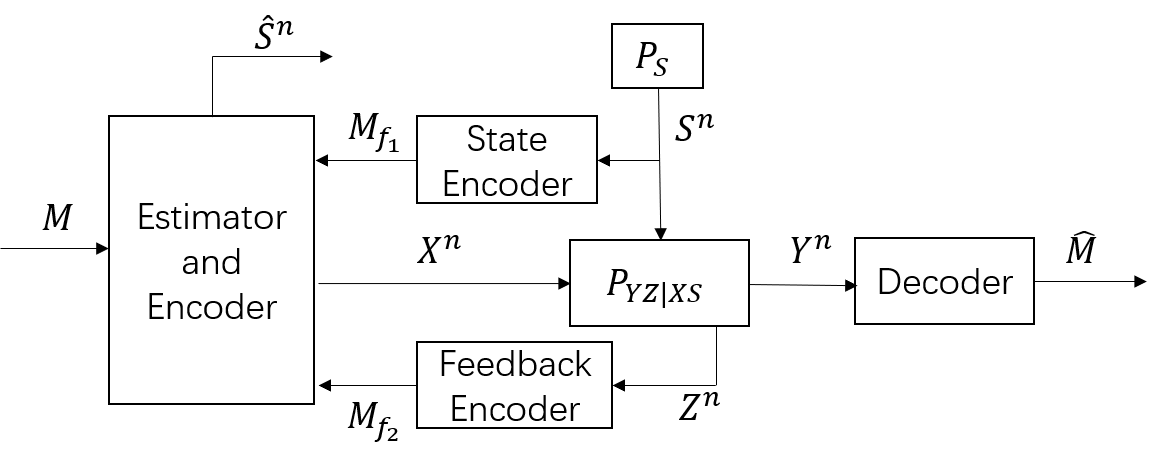}
    \caption{The rate-limited ISAC System: The state encoder observes the channel state noncausally and sends the sender a message $M_{f_1}$. The feedback encoder observes a feedback signal $Z^n$ and sends the sender a message $M_{f_2}$.}
    \label{fig:rate-limited isac}
\end{figure}

As depicted in Fig. \ref{fig:rate-limited isac}, the model considered in this paper is a state-dependent point-to-point discrete memoryless channel with an additional sender-side estimation of the channel state. The state encoder and the feedback encoder capture the channel state and the feedback signal, respectively, and produce the lossy descriptions $M_{f_1}$ and $M_{f_2}$. After receiving these lossy descriptions, the sender-side estimator estimates the channel state sequence and produces $\hat{S}^n$. On the other hand, normal message communication proceeds between the sender and receiver. In the following, we define the code for this model.
\begin{definition}
    An $(n,R,R_{f_1},R_{f_2})$ code for the ISAC system with rate-limited helpers consists of 
    \begin{itemize}
        \item a message set $\mathcal{M}=[1:2^{nR}]$,
        \item a state message set $\mathcal{M}_{f_1} = [1:2^{nR_{f_1}}],$
        \item a feedback signal message set $\mathcal{M}_{f_2} = [1:2^{nR_{f_2}}],$
        \item a state encoder: $f_1:\mathcal{S}^n \to \mathcal{M}_{f_1}$,
        \item a feedback signal encoder $f_2:\mathcal{Z}^n\to \mathcal{M}_{f_2}$,
        \item a message encoder $f: \mathcal{M}\times\mathcal{M}_{f_1}\to \mathcal{X}^n$,
        \item a decoder $g:\mathcal{Y}^n\to\mathcal{M}$ and
        \item a state estimator $h:\mathcal{M}\times \mathcal{M}_{f_1} \times \mathcal{M}_{f_2} \times \mathcal{X}^n  \to \hat{S}^n.$
    \end{itemize}
\end{definition}
The quality of the estimation is measured by the expectation of the distortion between the estimated sequence and the original sequence
\begin{align}
    \mathbb{E}\left[ d(S^n,\hat{S}^n) \right] = \frac{1}{n}\sum_{i=1}^n\mathbb{E}[d(S_i,\hat{S}_i)],
\end{align}
where $d:\mathcal{S}\times\hat{\mathcal{S}}\to [0,+\infty).$
\begin{definition}
    A rate-compression-distortion tuple $(R,R_{f_1},R_{f_2},D)$ is achievable if for any $\epsilon>0$, there exists a sufficiently large $N$ such that for any $n > N$ there exists a code $(n,R,R_{f_1},R_{f_2})$ such that
    \begin{align}
        Pr\{\hat{M}\neq M\} \leq \epsilon,\\
        \mathbb{E}\left[ d(S^n,\hat{S}^n) \right] \leq D.
    \end{align}
    The capacity-compression-distortion tradeoff $C(D)$ is the set of all tuples $(R,R_{f_1},R_{f_2})$ such that $(R,R_{f_1},R_{f_2},D)$ is achievable.
\end{definition}
\subsection{Main Results}
Define a set of rate tuples $(R,R_{f_1},R_{f_2})$ with auxiliary random variables $(Q,V_{11},V_{12},V_{2},U)$ taking values from finite alphabets $(\mathcal{Q},\mathcal{V}_{11},\mathcal{V}_{12},\mathcal{V}_2,\mathcal{U})$ such that
    \begin{equation}\label{def: achievable region}
        \mathcal{R}^{in}(D) = \left\{
            \begin{aligned}
                & R \leq I(U;Y|Q)-I(U;V_{11}|Q),\\
                & R_{f_{1}} \geq I(V_{11};S|Q) + I(V_{12};S,Z|U,V_{11},V_2,Q),\\
                & R_{f_2} \geq I(V_2;S,Z|U,V_{11},V_{12},Q),\\
                & R_{f_{1}} + R_{f_2} \geq I(V_{11};S|Q) + I(V_{12},V_2;S,Z|U,V_{11},Q),
            \end{aligned}
        \right.
    \end{equation}
     and the joint distribution of the random variables satisfies
    \begin{align}
        \label{def: joint distribution}P_SP_QP_{V_{11}V_{12}|SQ}P_{U|V_{11}Q}P_{X|UV_{11}Q}P_{YZ|XS}P_{V_2|ZQ},
    \end{align}
     and there exists a deterministic function $h:\mathcal{Q}\times \mathcal{U}\times \mathcal{X}\times\mathcal{V}_{12}\times\mathcal{V}_{2}\to \hat{\mathcal{S}}$ such that
    \begin{align}
        \mathbb{E}\left[ d(S,h(Q,U,X,V_{12},V_{2}))  \right]\leq D.
    \end{align}
     The sizes of the auxiliary variable alphabets satisfy
    \begin{align}
        &|\mathcal{Q}| \leq 5,\\
        &|\mathcal{V}_{11}| \leq |\mathcal{S}||\mathcal{X}|+1,\\
        &|\mathcal{V}_{12}| \leq |\mathcal{S}||\mathcal{X}|(|\mathcal{S}||\mathcal{X}|+1)+4,\\
        &|\mathcal{U}| \leq |\mathcal{S}||\mathcal{X}|(|\mathcal{S}||\mathcal{X}|+1)+1,\\
        &|\mathcal{V}_{2}| \leq |\mathcal{Z}|+4.
    \end{align}

\begin{theorem}\label{the: inner bound}
    The capacity-distortion-compression region of the ISAC with rate-limited helpers satisfies
    \begin{align}
        \mathcal{R}^{in}(D) \subseteq C(D).
    \end{align}
\end{theorem}
In the following paragraphs, we first give a brief explanation of the roles of those auxiliary random variables in the region \eqref{def: achievable region}. Then, a sketch of the proof is provided.

\emph{Discussion of auxiliary random variables: }The random variable $Q$ is a time-sharing random variable. The rates on $R_{f_1}$ and $R_{f_2}$ show a tradeoff between the compression rate, communication rate, and distortion constraint. The auxiliary random variables $V_{11}$ and $V_{12}$ represent the use of the channel state information in noncausal and causal ways, respectively. The combination of them $(V_{11},V_{12})$ forms a lossy description of the channel state. In the noncausal use, the state encoder sends the description $V_{11}$ at the beginning of the transmission block. The message encoder then uses this lossy description to encode the message as Heegard and El Gamal's coding scheme \cite{heegard1983capacity}. Due to the feedback signal encoder observing the channel feedback in a strictly causal manner, in this case, the encoding-decoding process of the state does not use any side information and, hence, requires a relatively high compression rate $I(V_{11};S)$. On the other hand, the state encoder can wait until the feedback signal encoder observes the entire feedback sequence. Then, they compress the state and feedback sequences in a distributed lossy compression way\cite{el2011network} and send $(V_{12},V_2)$. Although this reduces the required compression rate, the message encoder does not have this additional side information $V_{12}$ in the encoding phase, and the communication rate cannot benefit from $V_{12}$ directly. The choice of $V_{11}$ and $V_{12}$ shows a tradeoff between the noncausal and causal use.

\emph{Sketch of the Achievability: } Here, we provide a sketch of the achievability proof. It is sufficient to consider the case that $|\mathcal{Q}|=1$ and the rest of the proof uses the time-sharing. The details can be found in Section \ref{sec: proof of the general inner bound}. At the beginning of each transmission block, the state encoder observes the channel states $S^n$. It generates and sends two indices $M_{f_{11}}$ and $M_{f_{12}}$, representing two sequences $V_{11}^n$ and $V_{12}^n$, to the sender. The feedback encoder waits until the end of the transmission block when it collects the whole feedback signal. Then it finds a lossy description $V_2^n$ and sends the index $M_{f_2}$. The index $M_{f_{11}}$ is sufficient for the sender to find $V_{11}^n$ while $V^n_{12}$ and $V_2^n$ can only be decoded when the sender has both $M_{f_{12}}$ and $M_{f_2}$. At the encoder side, it decodes $M_{f_{11}}$ and finds $V^n_{11}$ once it receives the message from the state encoder. To this end, a constraint $R_{f_{11}}\geq I(V_{11};S)$ is required. To transmit a message, the encoder adopts Heegard and El Gamal's coding scheme\cite{heegard1983capacity} with $V_{11}^n$ being the rate-limited side information. Hence, a communication rate
\begin{align}
    R \leq I(U;Y) - I(U;V_{11})
\end{align}
is achieved. However, the encoder is not able to decode $V_{12}^n$ based solely on $M_{f_2}$ as $M_{f_{12}}$ and $M_{f_2}$ are descriptions when $(S^n,Z^n)$ are compressed jointly. The sender has to wait until the end of the block to receive $M_{f_2}$ then decodes $(V^n_{12},V^n_2)$ jointly. The constraints on $R_{f_{12}}$ and $R_{f_2}$ satisfy
\begin{align}
    & R_{f_{12}} \geq I(V_{12};S,Z|U,V_{11},V_2),\\
    & R_{f_2} \geq I(V_2;S,Z|U,V_{11},V_{12}),\\
    & R_{f_{12}} + R_{f_2} \geq I(V_{12},V_2;S,Z|U,V_{11}).
\end{align}
The proof is completed by applying the Fourier-Motzkin Elimination to all the constraints and the time-sharing argument.

\begin{remark}
    The region in Theorem \ref{the: inner bound} can be easily extended to a multi-letter region by considering the n-fold product channel. However, it is still not clear if the multi-letter region outperforms the single-letter region due to the existence of the auxiliary random variables. Similarly, the achievable regions in the following sections can also be extended to multi-letter regions. For those cases, single-letter outer bounds are also provided, and one can find that the inner and outer bounds only differ from a conditional distribution. Hence, the gain of the multi-letter regions, if a gain can be achieved at all, is relatively tiny.
\end{remark}
The region in Theorem \ref{the: inner bound} is an inner bound and generally not tight. 
In the rest of this section, we consider the case that the state sequence is available at the decoder side and the feedback signal is perfect, i.e. $Z=Y$. The feedback helper in the following two subsections are rate-unlimited and rate-limited, respectively.
\subsection{Rate-limited ISAC with Perfect Feedback}\label{sec: rate limited isac with perfect feedback}
    Let region $\mathcal{R}^{in}_{PF}$ be the set of rate pairs $(R,R_{f_1})$ such that
    \begin{align}\label{def: inner bound perfect feedback}
        \mathcal{R}^{in}_{PF}(D) = \left\{ 
            \begin{aligned}
                & R \leq I(X;Y|V_{11},S),\\
                & R_{f_{1}} \geq I(V_{11};S) + I(V_{12};S|X,Y,V_{11}),
            \end{aligned}
        \right. 
    \end{align}
    with the joint distribution $P_SP_{V_{11}V_{12}|S}P_{X|V_{11}}P_{Y|XS}$
    such that $\mathbb{E}\left[ d(S,h(X,Y,V_{12})) \right]\leq D$, where $h:\mathcal{X}\times\mathcal{Y}\times\mathcal{V}_{12}\to\hat{\mathcal{S}}$ is a deterministic function. The sizes of the auxiliary variable alphabets satisfy
    \begin{align}
        \label{ine: cardinality bound of perfect feedback 1}&|\mathcal{V}_{11}|\leq |\mathcal{S}||\mathcal{X}|+1,\\
        \label{ine: cardinality bound of perfect feedback 2}&|\mathcal{V}_{12}| \leq |\mathcal{S}|(|\mathcal{S}||\mathcal{X}|+1) + 1.
    \end{align}
    Further, define the region $\mathcal{R}^{out}_{PF}(D)$ having the same expression as \eqref{def: inner bound perfect feedback} but with the joint distribution
    \begin{align}
        P_SP_{V_{11}V_{12}|S}P_{X|V_{11}V_{12}}P_{Y|XS}
    \end{align}
    \begin{corollary}\label{coro: rate unlimited feedback}
        The capacity-compression-distortion region of the ISAC with an informed decoder and a rate-unlimited feedback signal helper satisfies
        \begin{align}
            \mathcal{R}^{in}_{PF}(D) \subseteq C_{PF}(D) \subseteq \mathcal{R}^{out}_{PF}(D).
        \end{align}
    \end{corollary}
    The achievability part combines the coding scheme in \cite{rosenzweig2005channels} and the compression part of the coding scheme for Theorem \ref{the: inner bound}. The auxiliary random variable $V_{11}$ is the common information between the encoder and the decoder due to the fact that the state sequence is known to the decoder. Since the feedback signal is perfect feedback and the feedback encoder is rate unlimited, we further set $Z=Y$ and $V_2=Y$ in Theorem \ref{the: inner bound}. This proves the achievability. The converse part is provided in Section \ref{sec: converse of rate unlimited feedback}.
    The difference between the inner and outer bounds is in the factorization of the joint pmf where we have the conditional distributions $P_{X|V_{11}}$ and $P_{X|V_{11}V_{12}}$, respectively. The inner and outer bounds meet when the state encoder is cognizant of the transmitted message, as we will discuss in Section \ref{sec: message cognizant case}.

    \subsection{Decomposable Channel with Perfect Feedback}\label{sec: decomposobal channel}
    For a more general case where the feedback encoder is rate-limited but still required to compress the signal losslessly, we consider a particular channel model such that the joint distribution $P(s,y) \neq 0$ does not hold for all $(s,y)\in\mathcal{S}\times\mathcal{Y}$. This happens when disjoint sets of channel states result in disjoint sets of channel output. We say that the joint distribution $P_{SY}$ is \emph{decomposable} in this case, and, therefore, the channel is \emph{decomposable}. For such a channel model, as shown by Witsenhausen in \cite{witsenhausen1975sequences}, the state sequence and the feedback sequence have a common component no matter what input distribution is selected. That is to say, there exist two deterministic functions $c_1:\mathcal{S}\to \mathcal{K}_1$ and $c_2:\mathcal{Y}\to\mathcal{K}_2$ such that
    \begin{align}
        c_1(S)=K_1,\; c_2(Y)=K_2,\; K_1=K_2=K,\;\;\;a.s..
    \end{align}
    
    For sources with common components, Berger-Tung type coding is sub-optimal since the encoders can agree on some specific quantization of $K^n$. Still, they compress this common sequence $k^n$ separately; see details in \cite{wagner2011distributed}. Similarly, for the ISAC with rate-limited helpers over a decomposable channel, the compressors can also agree on a sequence once they observe the sources.
     Consider an ISAC system with rate-limited helpers over a decomposable channel. If the feedback signal encoder is rate-limited and compresses the signal losslessly, the following result holds.
    Define a set of rate pairs $(R,R_{f_1},R_{f_2})$ such that
         \begin{align}\label{def: inner bound of common components 1}
            \mathcal{R}^{in,1}_d(D) = \left\{  
                \begin{aligned}
                &R \leq I(X;Y|S,V_{11},T),\\
                &R_{f_{1}} \geq I(V_{11};S) +I(T;S|V_{11}) + I(V_{12};S|X,Y,V_{11},T),\\
                &R_{f_2} \geq H(Y|X,V_{11},V_{12},T),\\
                &R_{f_1}+R_{f_2} \geq I(V_{11};S) +I(T;S|V_{11})+I(V_{12};S|X,Y,V_{11},T)+H(Y|X,V_{11},T),
                \end{aligned}
            \right.
        \end{align}
        and the random variables are defined by the joint distribution
        \begin{align}\label{def: common component inner bound joint distribution 1}
          P_{S}P_{V_{11}|S}P_{T|KV_{11}}P_{V_{12}|STV_{11}}P_{X|V_{11}}P_{Y|XS}.
        \end{align}
        and there exists a deterministic function $h:\mathcal{T}\times\mathcal{V}_{12}\times\mathcal{X}\times\mathcal{Y}\to\hat{\mathcal{S}}$ such that
        \begin{align}\label{def: common component inner bound distortion constraint}
            \mathbb{E}\left[ d(S,h(T,V_{12},X,Y)) \right] \leq D.
        \end{align}
    Define another set of rate pairs $(R,R_{f_1},R_{f_2})$ such that
        \begin{align}\label{def: inner bound of common components 2}
            \mathcal{R}^{in,2}_d(D) = \left\{  
                \begin{aligned}
                &R \leq I(X;Y|S,V_{11}),\\
                &R_{f_{1}} \geq I(V_{11};S)+ I(V_{12};S|X,Y,V_{11},T),\\
                &R_{f_2} \geq I(T;Y|V_{11}) + H(Y|X,V_{11},V_{12},T),\\
                &R_{f_1}+R_{f_2} \geq I(V_{11};S)+I(T;Y|V_{11}) + I(V_{12};S|X,Y,V_{11},T) + H(Y|X,V_{11},T)
                \end{aligned}
            \right.
        \end{align}
        and the random variables are defined by the joint distribution
        \begin{align}\label{def: common component inner bound joint distribution 2}
          P_{S}P_{V_{11}|S}P_{T|K}P_{V_{12}|STV_{11}}P_{X|V_{11}}P_{Y|XS}.
        \end{align}
    with the deterministic function $h:\mathcal{T}\times\mathcal{V}_{12}\times\mathcal{X}\times\mathcal{Y}\to\hat{\mathcal{S}}$ defined as \eqref{def: common component inner bound distortion constraint}.
    Further define region $\mathcal{R}_d^{out}(D)$ by
         \begin{align}\label{def: outer bound of common components}
            \mathcal{R}^{out}_d(D) = \left\{  
                \begin{aligned}
                &R \leq I(X;Y|S,V_{11}),\\
                &R_{f_{1}} \geq I(V_{11};S) + I(V_{12};S|X,Y,V_{11},T),\\
                &R_{f_2} \geq H(Y|X,V_{11},V_{12},T),\\
                &R_{f_1} + R_{f_2} \geq I(V_{11};S)+I(T;S|V_{11},X)+I(V_{12};S|T,X,Y,V_{11}) + H(Y|T,V_{11},X)
                \end{aligned}
            \right.
        \end{align}
        and the random variables are defined by the joint distribution
        \begin{align}\label{def: common component inner bound joint distribution}
            P_SP_{V_{11}|S}P_{X|V_{11}}P_{T|SV_{11}X}P_{V_{12}|STV_{11}X}P_{Y|XS},
        \end{align}
   with a function $h$ such that \eqref{def: common component inner bound distortion constraint} holds. Let $\mathcal{R}^{in}_d(D)$ be the convex hull of 
   \begin{align}
        \mathcal{R}^{in,1}_d(D) \cup \mathcal{R}^{in,2}_d(D)
   \end{align}
    \begin{theorem}\label{the: common component inner and outer bounds}
        For the ISAC over decomposable channels with an informed decoder and losslessly compressed feedback,
        the capacity-distortion region satisfies
        \begin{align}
            \mathcal{R}^{in}_d(D)\subseteq C(D) \subseteq \mathcal{R}^{out}_d(D).
        \end{align}
    \end{theorem}
    The regions $\mathcal{R}^{in,1}_d(D)$ and $\mathcal{R}^{in,2}_d(D)$ correspond to the cases in which the common component $T$ is either transmitted by the state encoder or the feedback encoder, respectively. Note that the joint distributions \eqref{def: common component inner bound joint distribution 1} and \eqref{def: common component inner bound joint distribution 2} are different depending on the choice of distributions $P_{T|K}$ and $P_{T|KV_{11}}$.
    For region $\mathcal{R}^{in,1}_d(D)$, the state encoder sends the lossy description of the common component, and the sum rate satisfies
    \begin{align}
        &I(V_{11};S) +I(T;S|V_{11})+I(V_{12};S|X,Y,V_{11},T)+H(Y|X,V_{11},T)\\
        &=I(V_{11};S) +I(T;S|V_{11}) + I(V_{12};S|X,V_{11},T)+H(Y|X,V_{11},V_{12},T).
    \end{align}
    The above equality indicates two corner points of the region:
    \begin{itemize}
        \item $(I(X;Y|S,V_{11}),I(V_{11};S)+I(T;S|V_{11})+I(V_{12};S|T,X,V_{11}),H(Y|T,X,V_{11},V_{12})),$
        \item $(I(X;Y|S,V_{11}),I(V_{11};S)+I(T;S|V_{11})+I(V_{12};S|T,X,V_{11},Y),H(Y|T,X,V_{11})),$
    \end{itemize}
    which are determined by different choices of the decoding orders:
    \begin{itemize}
        \item $V_{11}-T (\text{by the state helper})-V_{12}-Y$,
        \item $V_{11}-T (\text{by the state helper})-Y-V_{12}$.
    \end{itemize}
    The region $\mathcal{R}^{in,2}_d(D)$ is the case in which the feedback encoder sends the description $T$. Although the message sender received $V_{11}$ at the beginning of the transmission block, it can only be used as decoder-side information for $T$ since the feedback encoder has no access to $V_{11}$ and the distribution of $T$ is restricted to $P_{T|K}$ instead of $P_{T|KV_{11}}$. It implies the following two corner points:
    \begin{itemize}
        \item $(I(X;Y|S,V_{11}),I(V_{11};S)+I(V_{12};S|T,X,V_{11},Y),I(T;Y|V_{11}) + H(Y|T,X,V_{11})),$
        \item $(I(X;Y|S,V_{11}),I(V_{11};S)+I(V_{12};S|T,X,V_{11}),I(T;Y|V_{11}) + H(Y|T,X,V_{11},V_{12})),$
    \end{itemize}
    which are determined by the decoding order
    \begin{itemize}
        \item $V_{11}-T (\text{by the feedback helper})-Y-V_{12}$,
        \item $V_{11}-T (\text{by the feedback helper})-V_{12}-Y$.
    \end{itemize}
    We give the coding scheme for the second set of corner points of the region $\mathcal{R}^{in,1}_d(D)$ in Section \ref{sec: inner bound proof of common components}. The converse is presented in Section \ref{sec: converse common component}.

    As we mentioned earlier, the two regions $\mathcal{R}^{in,1}_d$ and $\mathcal{R}^{in,2}_d$ comes from the fact that the lossy description $V_{11}$ is not available at the feedback encoder side. This limitation of the helper results in two different factorizations of the joint distributions, and the discrepancy disappears when the receiver is also the feedback encoder (which is equivalent to the case that the state sequence is also available at the feedback helper side).   
    In this case, we no longer need the decomposable assumption since the source at the state encoder side $S$ is always a deterministic function of the source on the receiver side (which is also the feedback encoder) $(S,Y)$. Hence, the receiver-side encoder case can be regarded as a special case of the channel with the common component since the state is a deterministic function of the receiver's observation. Both $V_{11}$ and $T$ are descriptions of the state sequences, and the receiver can also help to send $T$ with $V_{11}$ being common side information since it has full information about the state sequence. 
    We call the feedback helper in this case a receiver-side helper (RH). In the following subsection, with the RH assumption, we consider a message-cognizant state encoder such that the capacity-compression-distortion region is achieved.
    \subsection{Capacity-achieving Case: Message-cognizant State Encoder}\label{sec: message cognizant case}
    When the receiver is also the feedback helper, the difference between the inner and outer bounds is due to the conditional distribution $P_{X|V_{11}}$ and $P_{X|V_{11}V_{12}}$ in the factorization of the joint pmf. In the proposed coding scheme, the codeword $X^n$ is only used as decoder side information for decoding $V_{12}$ and $V_2$ and has the Markov chain relation $V_{12}-V_{11}-X$. In contrast, in $\mathcal{R}^{out}_d(D)$, it serves as the common information at both the message transmitter side and the state encoder side.
    The coding scheme for compressing $T^n$ is a simple point-to-point lossy compression with common side information, and the compressing for $V^n_{12}$ in Section \ref{sec: inner bound proof of common components} uses a Wyner-Ziv compression and hence requires a Markov chain relation $X-(S,V_{11})-V_{12}$. 
     
     In the following, we consider a message-cognizant state encoder such that it is informed of the decoding result by the decoder at the end of each transmission block. This allows the state encoder to use the information of the message codeword at the end of each transmission block. In this case, we have the following capacity result.
     \begin{corollary}\label{coro: capacity of message cognizant state encoder}
         For an ISAC system defined in Theorem \ref{the: common component inner and outer bounds} with a message-cognizant state encoder and a receiver-side feedback encoder, the capacity-compression-distortion region satisfies 
         \begin{align}
             C_{MC-RH}(D) = \mathcal{R}^{out}_d(D),
         \end{align}
        where the subscript `MC-RH' represents message-cognizant and receiver-side helper. 
     \end{corollary}
     
    The difference between our message-cognizant state encoder and that in \cite{lapidoth2023state} and \cite{lapidoth2023state2} is that in this paper, the state encoder does not know the message prior to the current block. Instead, it is informed at the end of the block after the decoder produces a guess of the transmitted message.
     
     To use this additional information at the state encoder side, we propose a block Markov coding scheme built upon the scheme in Section \ref{sec: inner bound proof of common components}. The sketch of the coding scheme is as follows.

    In each block, the codebook generations of $\{v^n_{11}\}$ and $\{x^n|n_{11}\}$ are exactly the same as that of Section \ref{sec: inner bound proof of common components}. For each pair of $(v^n_{11}(n_{11}),x^n(n_{11},m))$, generate the common component codebook $\{t^n|n_{11},m\}$ with codebook size $N_{0}=\exp\{n(I(T;S|V_{11},X))\}$ and codebook $\{v^n_{12}|n_{11},m,n_{0}\}$ with size $\exp\{n(R_{f_{12}}+\epsilon)\}$. In the following, we use notation $S^n(b)$ to indicate that the state sequence is in Block $b$. The encoding process is as follows.
    \begin{itemize}
        \item  In Block 1, the state helper observes the state $S^n(1)$. It sends the first index $\phi_{f_{11}}(S^n(1))$ and sets the second and third indices $(\phi_0(S^n(1)),\phi_{f_{12}}(S^n(1)))$ as constants.
        \item In Block $b\in[2:B-1]$, the state helper observes the state $S^n(b)$. It sends the first index $\phi_{f_{11}}(S^n(b))$ and uses the second and third indices $(\phi_0(S^n(b-1)),\phi_{f_{12}}(S^n(b-1)))$ with $X^n(b-1)$ being the additional common information to describe the state sequence of the previous block $b-1$. Note that the codeword is determined by $n_{11}$ and $\hat{m}$; hence, the encoding error occurs when the decoding result is wrong.
        \item In Block B, the state helper observes the state $S^n(B)$, sets the first index to be a constant, and uses the second and third indices $(\phi_0(S^n(B-1)),\phi_{f_{12}}(S^n(B-1)))$ with $X^n(B-1)$ being common information to describe the state sequence of the previous block $B-1$.
    \end{itemize}  
In this block Markov coding scheme, only $V_{11}$ is decoded in the current transmission block $b$ before the sender side encoding to facilitate the communication, while $T$ and $V_{12}$ are decoded in the next transmission block $b+1$ with the codeword $X^n(b)$ in block $b$ being common side information. 
Similarly, with a message-cognizant state encoder, the outer bound in Corollary \ref{coro: rate unlimited feedback} becomes tight.
\begin{corollary}\label{coro: message cognizant unlimited feedback}
    For an ISAC system defined in Section \ref{sec: rate limited isac with perfect feedback} with a message-cognizant state encoder, the capacity-compression-distortion region satisfies
    \begin{align}
        C_{MC-PF}=\mathcal{R}^{out}_{PF}(D).
    \end{align}
\end{corollary}
In the following, we apply the results we have so far to different communication scenarios and show some numerical results.
\section{examples and applications}\label{sec: examples and applications}
\subsection{Z-channel with Random Parameters}
\begin{figure}
    \centering
    \includegraphics[width=0.5\linewidth]{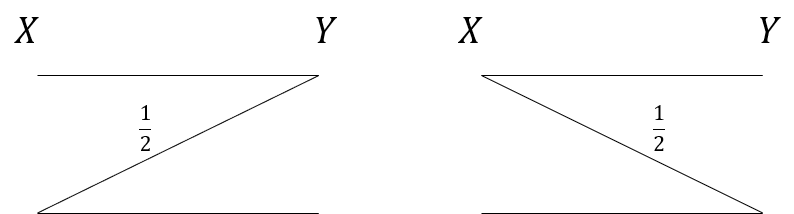}
    \caption{Z-channel with random parameters $S$}
    \label{fig:z channel example}
\end{figure}

In this subsection, we investigate the impact of the choice of $V_{11}$ on the communication and compression rates under a fixed distortion level $D$. Consider a two-state binary channel with rate-unlimited feedback such that the transition probability under each state forms a Z-channel, as depicted in Fig.\ref{fig:z channel example} and $S\sim Bernoulli(\frac{1}{2})$. 
We assume the state is available at the decoder side and the state helper is message-cognizant. Then, we can apply the result in Corollary \ref{coro: message cognizant unlimited feedback} to this model. To get an inner bound of the capacity-compression-distortion region for a given distortion level $0\leq D \leq \frac{1}{2}$, we construct the auxiliary random variables as follows:
\begin{align}
    &V_{11} = S\oplus N_1 \oplus N_2,\;\;N_1\sim Bernoulli(D), N_2 \sim Bernoulli(D'),\\
    &V_{12} = X\oplus S\oplus N_1,\\
    &\hat{S} = X \oplus V_{12}.
\end{align}
where $\oplus$ is the modulo 2 addition.
We further set the input distribution as
\begin{align}
    X\sim Bernoulli(\alpha)\;\;\text{if $V_{11}=0$},\\
    X\sim Bernoulli(\beta)\;\;\text{if $V_{11}=1$},
\end{align}
where $\alpha=\frac{2}{5},\beta=\frac{3}{5}$.
By simple algebra computation, $X\sim Bernoulli(\frac{2}{5})$ is the capacity-achieving input distribution of the Z-channel when $S=0$, and $X\sim Bernoulli(\frac{2}{5})$ is the capacity-achieving input distribution of the Z-channel when $S=1$. With this coding strategy, the encoder always assumes that $V_{11}=S$ in the encoding phase.

Substituting the random variables into Corollary \ref{coro: message cognizant unlimited feedback} gives the coding rate and compression rate \begin{align}
     &I(X;Y|S,V_{11})\\
    & = \frac{1-D*D'}{2}(H_b(1-\frac{\alpha}{2})+H_b(\frac{1-\beta}{2}))+\frac{D*D'}{2}(H_b(1-\frac{\beta}{2})+H_b(\frac{1-\alpha}{2}) ) \\
    &- \frac{(1-D*D')\alpha+(D*D')\beta}{2} - \frac{D*D'(1-\alpha)+(1-D*D')(1-\beta)}{2},\\
    &I(S;V_{11}) = 1 - H(D*D'),\\
    &I(S;V_{12}|X,V_{11}) = H_b(D*D') - H_b(D),\\
    &H(Y|X,V_{11}) = \frac{1}{2}(1-\alpha+\beta) H_b(\frac{D*D'}{2}) + \frac{1}{2}(1-\beta+\alpha)H_b(\frac{1+D*D'}{2}),\\
    &H(Y|X,V_{11},V_{12}) = \frac{1}{2}(1-\alpha+\beta+2\alpha D' -2\beta D')H_b(\frac{D}{2})+\frac{1}{2}(1+\alpha-\beta-2\alpha D' + 2\beta D')H_b(\frac{1+D}{2})
\end{align}
\begin{figure}
    \centering
    \includegraphics[width=0.5\linewidth]{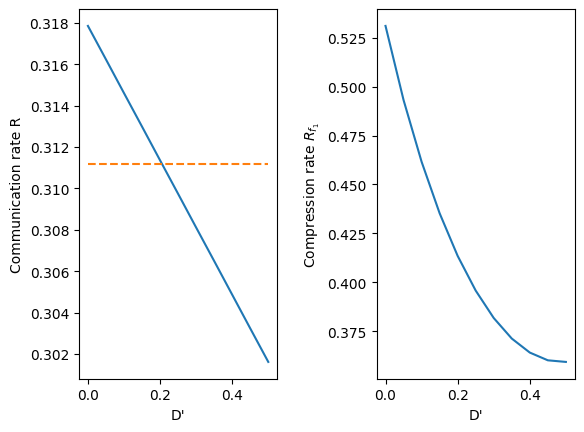}
    \caption{The impact of $D'$ on the communication rate $R$ and the compression rate $R_{f_1}$. The dotted line is the capacity of the channel when no side information is available at the encoder side, and there is no distortion constraint.}
    \label{fig:z channel rate compression}
\end{figure}
Since the auxiliary random variable $V_{11}=S\oplus N_1 \oplus N_2=S \oplus \widetilde{N}, \widetilde{N}\sim Bermoulli(D*D')$, we can consider it as a lossy description of the state $S$, and its description quality depends on $D*D'$, where $D'$ is the refinement parameter. When $D'=\frac{1}{2}$, we have $D*D'=\frac{1}{2}$ and $V_{11}$ is a uniformly distributed random variable independent of $S.$ In this case, the state helper uses the channel state information causally, and the message encoder does not benefit from observing $V_{11}$. As shown in Fig. \ref{fig:z channel rate compression}, both the communication and compression rates are low. When $D'$ gets closer to $0$, $V_{11}$ becomes a more precise description of $S$, and the communication rate increases. The dotted line in Fig.\ref{fig:z channel rate compression} is the capacity of this channel model achieved by uniform input distribution when no side information is available at the encoder and no distortion constraint is imposed. It indicates that when $D'$ is smaller than a certain number, the encoder benefits from having $V_{11}$ and achieves a higher communication rate than the no-side-information case.  In the meantime, a small $D'$ means that more information from the channel state is used noncausally, resulting in a higher compression rate.

\begin{figure}
    \centering
    \includegraphics[width=0.5\linewidth]{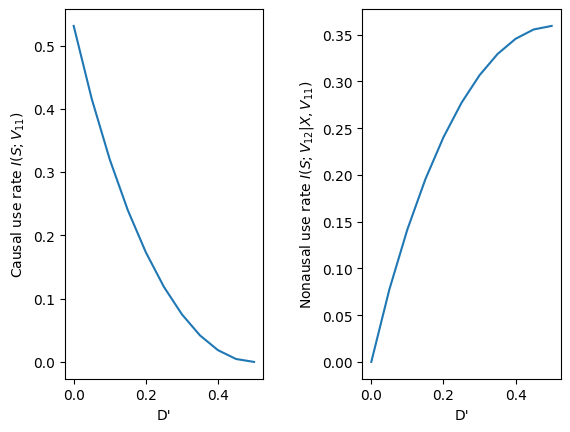}
    \caption{The tradeoff between $I(V_{11};S)$ and $I(V_{12};S|X,Y,V_{11})$ depending on different values of $D'$.}
    \label{fig:z channel causal noncausal}
\end{figure}

From the above discussion, one can consider $V_{11}$ as a coarse description of $S$ with rate $R_{f_{11}}=I(S;V_{11})$, and is then refined by receiving $V_{12}$ with rate $R_{f_{12}}=I(V_{12};S|X,Y,V_{11})$. We are also interested in the tradeoff between $V_{11}$ and $V_{12}$. As shown in Fig.\ref{fig:z channel causal noncausal}, when $D'=\frac{1}{2}$, $V_{11}$ is an independent random variable and does not offer any information about $S$. In this case, we have $R_{f_{11}}=I(V_{11};S)=0$ and $R_{f_{12}}$ reaches its maximum. All the information about $S$ is compressed, with the feedback $Y^n$ being the side information. On the other hand, when $D'=0$, $V_{11}$ is a description of $S$ such that the distortion constraint is satisfied. There is no need to further construct $V_{12}$ and hence $R_{f_{12}}=0$ and $R_{f_{11}}$ achieves its maximum. It should also be noted that the maximal value of $I(V_{12};S|X,Y,V_{11})$ is smaller than the maximal value of $I(V_{11};S)$ since it is the case that the feedback signal $Y^n$ is also used as side information for the compression.

In the following subsection, we completely characterize the capacity-compression-distortion region when successive refinement constraints are imposed on the estimation.

\subsection{Successive Refinement of the State}
In the previous results, the state encoder has great freedom to choose the distribution of $V_{11}$ depending on whether the system wants a higher communication rate or a lower compression rate. In addition to that, the decoding of $V_{11}$ and $(T,V_{12})$ can be considered as a successive refinement process of the state sequence. 

 Suppose that the estimator is required to reproduce the state sequence at the beginning of the transmission with distortion level $D_1$. At the end of the current block, it reconstructs a refined state sequence with distortion level $D_2$ such that $D_1>D_2$. Such a successive refinement is possible since the state encoder sends its lossy description to the encoder at the beginning of the block, which can be used to produce a rough estimation of the state sequence. Once the transmission is completed, the estimator receives a feedback signal (or its lossy version) and is able to refine the estimated sequence. 
\begin{corollary}
    For the ISAC model defined in Corollary \ref{coro: message cognizant unlimited feedback} with successive refinement constraint, the capacity-distortion region satisfies
    \begin{align}\label{def: capacity of successive refinment constraint}
        C_{SR-MC-PF}(D) = \left\{ 
            \begin{aligned}
                & R \leq I(X;Y|V_{11},S),\\
                & R_{f_{1}} \geq I(V_{11};S) + I(V_{12};S|X,Y,V_{11}),
            \end{aligned}
        \right. 
    \end{align}
    with the joint distribution 
    \begin{align}
        P_SP_{V_{11}V_{12}|S}P_{X|V_{11}V_{12}}P_{Y|XS}
    \end{align}
    such that $\mathbb{E}\left[ d(S,h_1(V_{11})) \right]\leq D_1,\mathbb{E}\left[ d(S,h(X,Y,V_{12})) \right]\leq D_2$, where $h_1:\mathcal{V}_{11}\to\hat{\mathcal{S}}$ and $h_2:\mathcal{X}\times\mathcal{Y}\times\mathcal{V}_{12}\to\hat{\mathcal{S}}$ are deterministic functions. The cardinality bounds of the auxiliary random variables are
    \begin{align}
        \label{ine: cardinality bound of perfect feedback and successive refinement 1}&|\mathcal{V}_{11}|\leq |\mathcal{S}||\mathcal{X}|+2,\\
        \label{ine: cardinality bound of perfect feedback and successive refinement 2}&|\mathcal{V}_{12}| \leq |\mathcal{S}|(|\mathcal{S}||\mathcal{X}|+1) + 1.
    \end{align}
\end{corollary}

    The achievability proof is the same as in Section \ref{sec: rate limited isac with perfect feedback} with an additional constraint $D_1$. For the converse, note that in Section \ref{sec: converse of rate unlimited feedback}, we set $V_{11}=(M_{f_1},Q)$. Let $\hat{S}^n_1$ be the state sequence produced in the first-stage estimation. Then, for the distortion constraint, we have
    \begin{align}
        D_1 &\geq \frac{1}{n}\sum_{i=1}^n \mathbb{E}[d(S_i,\hat{S}_{1,i})]\\
        &\overset{(a)}{=}\frac{1}{n}\sum_{i=1}^n \mathbb{E}[d(S_i,h_1(V_{11,i}))],
    \end{align}
where $(a)$ follows by the fact that $\hat{S}_i$ can be determined by the lossy description $M_{f_1}$ and the index $i$, and the definition of $V_{11,i}=(M_{f_1},i).$

\subsection{Binary Channel with Fading}
Consider the following binary fading channel
\begin{align}
    Y=(X\cdot S) \oplus N,
\end{align}
where $S$ is Bernoulli$(\frac{1}{2})$, $N$ is Bernoulli$(q)$. Let $[a]^+=\max\{a,0\}.$
\begin{corollary}\label{coro: binary channel with fading and successive refinement}
    An outer bound of the capacity-distortion-compression region for the binary fading channel with successive refinement constraints $D_1$ and $D_2$ is
    \begin{align}\label{region: binary channel with fading example}
        C_{SR-PF}(D_1,D_2) = \bigcup_{\alpha\in[0:1]} \left\{ 
            \begin{aligned}
                & R\leq \frac{1}{2}H_b(\alpha * q)  - \frac{1}{2}H_b(q),\\
                & R_{f_{1}} \geq [1+\alpha H_b(q) - \alpha H_b(D_1*q) - H_b(D_2)]^+,
            \end{aligned}
        \right. 
    \end{align}
    An inner bound of the region has the same expression as \eqref{region: binary channel with fading example} but with a more strict constraint on $\alpha$: $\alpha\in[0,1]\;s.t.\; R_{f_1}>0.$
\end{corollary}
\emph{Converse:}
We start with the bound on the communication rate $R$.
\begin{align}
    I(X;Y|V_{11},S) &= H(Y|V_{11},S) - H(Y|X,S)\\
    &=H(X\cdot S \oplus N|V_{11},S) - H_b(N)\\
    &=P_S(0)H_b(N)+P_S(1)H(X\oplus N|V_{11},S=1) - H_b(N)\\
    &\leq P_S(1)H(X\oplus N) - P_S(1)H_b(N)\\
    &=\frac{1}{2}H_b(\alpha*q) - \frac{1}{2}H_b(q),
\end{align}
where $\alpha\in[0,1]$, $p*q=p(1-q)+(1-p)q,H_b(\cdot)$ is the binary entropy function.

For the compression rate $R_f$, it follows that
\begin{align}
    &I(V_{11};S) + I(V_{12};S|X,Y,V_{11})\\
    &=H(S)+ H(Y|X,S) - H(Y|X,V_{11}) - H(S|X,Y,V_{11},V_{12})\\
    &\overset{(a)}{=}H(S)+ H(Y|X,S) - H(X\cdot S \oplus N|X,V_{11}) - H(S|X,Y,V_{11},V_{12},\hat{S}_2)\\
    &\geq H(S)+ H(Y|X,S) - H(X\cdot S \oplus N|X,V_{11}) - H(S\oplus \hat{S}_2)\\
    &\label{eq: binary fading example 1}=H(S)+ H(Y|X,S) - H(X\cdot S \oplus N|X,V_{11}) - H_b(D_2)
\end{align}
where $(a)$ follows by the fact that $\hat{S}_2$ is determined by $(X,Y,V_{12})$. To bound $H(X\cdot S \oplus N|X,V_{11})$, we have
\begin{align}
    &H(X\cdot S \oplus N|X,V_{11})\\
    \label{eq: binary fading example 2}&=P_X(0)H_b(N) + P_X(1)H_b(S \oplus N|V_{11},X=1)\\
    &\leq P_X(0)H_b(N) + P_X(1)H_b(S \oplus \hat{S}_1 \oplus N).
\end{align}
Plugging it back into \eqref{eq: binary fading example 1} gives
\begin{align}
    R_{f_1}&\geq H_b(\frac{1}{2}) + H_b(q)- P_X(0)H_b(N)-P_X(1)H_b(S \oplus \hat{S}_1 \oplus N) -H_b(D_2)\\
    &=1 + P_X(1)H_b(q) - P_X(1)H_b(D_1*q)-H_b(D_2)\\
    &=1 + \alpha H_b(q) - \alpha H_b(D_1*q)-H_b(D_2).
\end{align}

\emph{Achievability: } For the achievability part, we construct the auxiliary random variable $V_{11}$ by a binary channel with crossover probability $D_1$, i.e. $S=V_{11}\oplus N_1$, where $N_1$ is Bernoulli($D_1$). The input distribution $P_{X|V_{11}}$ is defined by $P(X=0|V_{11}=0)=1-\alpha,P(X=0|V_{11}=1)=1-\alpha$.
It follows that 
\begin{align}
    I(X;Y|V_{11},S) &= H(Y|V_{11},S) - H(Y|X,S)\\
    &=P_S(0)H_b(N)+P_S(1)H(X\oplus N|V_{11},S=1) - H_b(N)\\
    &= P_S(1)H(X\oplus N|V_{11},S=1) - P_S(1)H_b(N)\\
    &=\frac{1}{2}D_1 H(X\oplus N|V_{11}=0) + \frac{1}{2}(1-D_1) H(X\oplus N|V_{11}=1) - \frac{1}{2}H_b(q)\\
    &=\frac{1}{2}H_b(\alpha * q)  - \frac{1}{2}H_b(q).
\end{align}
For the compression rate, setting $V_{12}= X+S+N_2$ gives
\begin{align}
    &I(V_{11};S) + I(V_{12};S|X,Y,V_{11})\\
    &=H(S)+ H(Y|X,S) - H(Y|X,V_{11}) - H(S|X,Y,V_{11},V_{12})\\
    &=H_b(\frac{1}{2}) + H_b(q) - P_X(0)H(N|X,V_{11}) - P_X(1)H(S\oplus N|X=1,V_{11})-H(S|X,Y,V_{11},V_{12})\\
    &=H_b(\frac{1}{2}) + \alpha H_b(q) -\alpha H(V_{11}\oplus N_1\oplus N|X=1,V_{11})-H(X\oplus N_2\oplus V_{12}|X,Y,V_{11},V_{12})\\
    &=1+\alpha H_b(q) - \alpha H_b(D_1*q) - H_b(D_2).
\end{align}
The state estimators are defined by $\hat{S}_1=V_{11}$ and $\hat{S}_2=X\oplus V_{12}$. The proof is complete.  

The gap between the inner and outer bounds is due to the case in which $1+\alpha H_b(q) - \alpha H_b(D_1*q) - H_b(D_2)$ is negative. To better understand the channel's performance in this case, we remove the successive refinement in the rest of this subsection and compare the resulting region with the one obtained by applying \cite[Theorem 1]{ahmadipour2022information} to the channel model.

To this end, we define a function
\begin{align}
    f(\alpha) = 1 - \alpha +\alpha H_b(q) - H_b(D)
\end{align}
and regions
\begin{align}
        \mathcal{R}_{SR-PF}^1(D) = \bigcup_{\substack{\alpha\in[0:1]:\\ f(\alpha)> 0}} \left\{ 
            \begin{aligned}
                & R\leq \frac{1}{2}H_b(\alpha * q)  - \frac{1}{2}H_b(q),\\
                & R_{f_{1}} \geq 1 - \alpha +\alpha H_b(q) - H_b(D),
            \end{aligned}
        \right. 
    \end{align}
\begin{align}
        \mathcal{R}_{SR-PF}^2(D) = \bigcup_{\substack{\alpha\in[0:1]:\\ \frac{1-\alpha}{2}+\alpha q \leq D,\\if \;f(\alpha)\leq 0}} \left\{ 
            \begin{aligned}
                & R\leq \frac{1}{2}H_b(\alpha * q)  - \frac{1}{2}H_b(q),\\
                & R_{f_{1}} = 0,
            \end{aligned}
        \right. 
    \end{align}
\begin{align}
        \mathcal{R}_{SR-PF}^3(D) = \bigcup_{\substack{\alpha\in[0:1]:\\ \frac{1-\alpha}{2}+\alpha q > D,\\if \;f(\alpha)\leq 0}} \left\{ 
            \begin{aligned}
                & R\leq \frac{1}{2}H_b(\alpha * q)  - \frac{1}{2}H_b(q),\\
                & R_{f_{1}} \geq (1-\alpha)(1-H_b(\frac{D-\alpha q}{1-\alpha})).
            \end{aligned}
        \right. 
    \end{align}
  Then we have the following corollary.
\begin{corollary}
    When $D\leq q$, the capacity-distortion-compression region for the binary fading channel is
    \begin{align}
        C_{SR-PF}(D) = \bigcup_{\substack{\alpha\in[0:1]:}} \left\{ 
            \begin{aligned}
                & R\leq \frac{1}{2}H_b(\alpha * q)  - \frac{1}{2}H_b(q),\\
                & R_{f_{1}} \geq 1 - \alpha +\alpha H_b(q) - H_b(D),
            \end{aligned}
        \right. 
    \end{align}
    When $q < D \leq \frac{1}{2}$, the capacity-compression-distortion region includes
    \begin{align}
        \mathcal{R}_{SR-PF}(D) = \bigcup_{i=1}^3 \mathcal{R}^i_{SR-PF}(D).
    \end{align}
\end{corollary}
The achievability of $C_{SR-PF}(D)$ and $\mathcal{R}_{SR-PF}^1(D)$ follows by setting $V_{11}$ independent of $S$ in the achievability part of Corollary \ref{coro: binary channel with fading and successive refinement} and the converse of $C_{SR-PF}(D)$ follows by removing it from the condition in the converse of Corollary \ref{coro: binary channel with fading and successive refinement}.
The difference between the inner and outer bounds in Corollary \ref{coro: binary channel with fading and successive refinement} is due to the positivity of the constraint of $R_{f_1}$. When $q < D < \frac{1}{2}$, note that
\begin{align}
    f(\alpha) = 1 - \alpha +\alpha H_b(q) - H_b(D)
\end{align}
may be negative since 
\begin{align}
    1 - \alpha +\alpha H_b(q)=(1-\alpha) H_b(\frac{1}{2}) + \alpha H_b(q)
\end{align}
is the convex combination of $H_b(\frac{1}{2})$ and $H_b(q)$.
\begin{figure}
    \centering
    \includegraphics[scale=0.75]{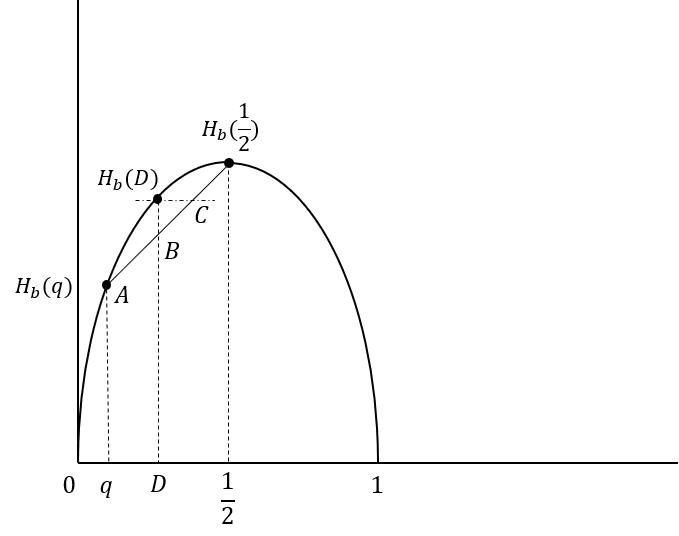}
    \caption{Relation between $H(q),H(D)$ and $1$.}
    \label{fig: binary fading example 0 compression rate}
\end{figure} 
As shown in Fig. \ref{fig: binary fading example 0 compression rate}, the function $f(\alpha)<0$ implies that we select some input distribution such that $1 - \alpha +\alpha H_b(q)$ falls into the line $\overline{AC}$. In the meantime, we set $R_{f_1}=0$ and estimate only based on the perfect feedback signal. In this case, we use the best estimator in \cite{ahmadipour2022information} as follows
\begin{align}
    \hat{s}(X,Y) = \left\{
        \begin{aligned}
            &0, \;\;X=0\\
            &Y, \;\;X=1.
        \end{aligned}
    \right.
\end{align}
The average distortion in this case is $\frac{1-\alpha}{2}+\alpha q$, which is the convex combination of points $\frac{1}{2}$ and $q$ again with $(1-\alpha,a)$ being the convex combination coefficients. However, the binary entropy function $H_b(\cdot)$ is a concave function, and therefore $f(\alpha)<0$ does not imply $\frac{1-\alpha}{2}+\alpha q < D$. In fact, there exists some choice of input distribution $\alpha$ such that $1 - \alpha +\alpha H_b(q)$ falls into the segment $\overline{BC}$. In this case, we have $f(\alpha)<0$, but the average distortion based only on the feedback signal is still greater than the constraint $D$. The achievable region in this case is $\mathcal{R}^2_{SR-PF}(D)$ and input distributions corresponding to the segment $\overline{BC}$ are not feasible. However, this constraint can be relaxed when there is a message-cognizant state helper. To see this, we set $V_{12}$ to be a constant when $X=1$ and $V_{12}=S+N_3$ when $X=0$, where $N_3\sim Bernoulli(\frac{D-\alpha q}{1-\alpha})$. For the estimator, it estimates $\hat{S}=Y$ when $X=1$ and $\hat{S}=V_{12}$ when $X=0$. The average distortion, in this case, is
\begin{align}
    \mathbb{E}[d(S,\hat{S})] = \alpha q + (1-\alpha) \frac{D-\alpha q}{1-\alpha} = D.
\end{align}
This gives the region $\mathcal{R}^3_{SR-PF}(D)$.
In fact, the region $\mathcal{R}^2_{SR-PF}(D)$ is the direct result of applying \cite[Theorem 1]{ahmadipour2022information} to our model with the distortion $\mathbb{E}[d(S,\hat{S})]=\frac{1-\alpha}{2}+\alpha q$ obtained by using the estimator defined in \cite[Lemma 1]{ahmadipour2022information}. The existence of the state helper relaxes the constraint on the input distribution. Hence, the auxiliary random variable $V_{12}$ is helpful for the message communication even if the message sender is not able to decode it when the sender encodes the message.

On the other hand, the minimal value of $\frac{1-\alpha}{2} + \alpha q$ is $q$ when $\alpha=1$. Therefore, any distortion $D<q$ cannot be achieved solely based on the feedback signal. The function $f(\alpha)$, in this case, is always positive, which indicates that when $D<q$, we always require an additional state helper to send some information with a positive rate to achieve a lower distortion, and we achieve the tight bound $C_{SR-PF}(D)$.

It should also be noted that when $q=\frac{1}{2}$, the noise is uniformly distributed, and no reliable communication is possible. The state helper compression rate in this case is $1-H_b(D)$, which is always positive for $D<\frac{1}{2}$ since the feedback signal is always uniformly distributed and is not helpful for the estimation. The estimator achieves the desired distortion level $D$ by the information from the state helper, which reduces the problem to the lossy source coding problem for a binary source with Bernoulli distribution $(\frac{1}{2},\frac{1}{2})$, and the result coincides with that in \cite[Theorem 10.3.1]{thomas2006elements}. The region is tight because a Bernoulli source under Hamming distortion is always successively refinable\cite[Example 13.2]{el2011network}, and constructing it into two steps does not hurt its optimality.

\subsection{Causal Helper}
In this subsection, we consider a special case in which the state encoder only observes the state in a causal manner. In this case, the capacity-distortion-compression region is as follows.
\begin{corollary}\label{coro: causal state encoder}
    For an ISAC model defined in Corollary \ref{coro: capacity of message cognizant state encoder} with a state encoder that observes the channel state causally, the capacity-compression-distortion region is the convex hull of
    \begin{align}
        \left\{
            \begin{aligned}
                &R \leq I(X;Y|S),\\
                &R_{f_1} \geq I(V;S|X,Y,T),\\
                &R_{f_2} \geq H(Y|X,V,T),\\
                &R_{f_1} + R_{f_2} \geq I(T,V;S|X) + H(Y|T,V,X).
            \end{aligned}
        \right.
    \end{align}
under the joint distribution 
\begin{align}
    P_SP_{X}P_{VT|SX}P_{Y|XS}
\end{align}
and a deterministic function $h: \mathcal{T}\times\mathcal{V}\times \mathcal{X} \times \mathcal{Y}\to \hat{\mathcal{S}}$ such that
\begin{align}
    \mathbb{E}[d(S,h(T,V,X,Y))]\leq D.
\end{align}
\end{corollary}
\begin{remark}
    Note that from the sum rate constraint, we have
    \begin{align}
        &I(T,V;S|X) + H(Y|T,V,X)\\
        &=I(T,V;S|X) + H(Y,T,V|X) - H(T,V|X)\\
        &=I(T,V;S|X) + H(Y|X) + H(T,V|X,Y) - H(T,V|X)\\
        &=I(T,V;S|X) + H(Y|X) - I(T,V;Y|X) = I(T,V;S|X,Y) + H(Y|X).
    \end{align}
    It indicates two corner points of the region corresponding to different decoding orders.
\end{remark}

The achievability follows by first setting $V_{11}=\emptyset$ and $V_{12}=V$ in Corollary \ref{coro: capacity of message cognizant state encoder}. Note that the resulting region is no longer convex in general. Hence, an additional time-sharing argument is necessary, and the achievable region in the corollary is the convex hull of the resulting region. The converse part is given in Section \ref{app: converse of causal state encoder}.

In the following, we study a binary example for the causal helper case. To this end, we consider a channel $Y=X\cdot S, S\sim Bernoulli(\frac{1}{2})$.
\begin{corollary}
The capacity-distortion-compression region for this model for some $D<\frac{1}{2}$ is the convex hull of
\begin{align}\label{coro: numerical binary fading with causal helper}
    \bigcup_{\substack{\Delta_1,\Delta_2 \in [0,\frac{1}{2}],\\\alpha\in[0,1]}} \left\{
        \begin{aligned}
            &R \leq \frac{1}{2}H_b(\alpha),\\
            &R_{f_1} \geq \left[ (1-\alpha)H_b(\Delta_1)-(1-\alpha) H_b(\min\{\frac{D}{1-\alpha},\frac{1}{2}\})\right]^+,\\
            &R_{f_2} \geq \alpha H_b(\Delta_2),\\
            &R_{f_1} + R_{f_2} \geq 1-(1-\alpha) H_b(\min\{\frac{D}{1-\alpha},\frac{1}{2}\}).
        \end{aligned}
    \right.
\end{align} 
\end{corollary}
\begin{remark}
    By choosing the input distribution $\alpha=1$, the rate of the state helper can be zero. In this case, the constraint on $R_{f_2}$ is inactive since we have the sum rate constraint
    \begin{align}
        R_{f_1}+R_{f_2}=R_{f_2}\geq 1,
    \end{align}
    which is the entropy of the state source. Note that $\alpha=1$ implies the sender sends the constant $X=1$, which follows $Y=S$. In this case, the feedback helper only needs to compress the state losslessly.

    On the other hand, by setting $\alpha=0$, the sender sends the constant $X=0$, and the channel output $Y=0$ is deterministic. Hence, the feedback helper does not need to say anything about the feedback signal, and the minimal rate is 0. In this case, the problem reduces to a lossy compression problem again, and the rate required at the state helper side is
    \begin{align}
        H(P_S)-H(D) = 1-H(D).
    \end{align}
\end{remark}
\begin{remark}
    This numerical example also shows that the region \eqref{coro: numerical binary fading with causal helper} of the causal state helper case without time-sharing is generally not convex. To show this, we set the rate $R_{f_1}$ to 0. In this case, the rate constraint on the feedback helper is 
    \begin{align}
        1-(1-\alpha) H_b(\min\{\frac{D}{1-\alpha},\frac{1}{2}\}).
    \end{align}
    By setting the input distribution $\alpha=0$, we achieve the rate-compression point $(0,0,1-H_b(D))$. On the other hand, for any $D < \Delta_1 \leq \frac{1}{2}$, there exists an $\hat{\alpha}$ such that 
    \begin{align}
        H_b(\Delta_1) = \frac{H_b(D)}{1-\hat{\alpha}}.
    \end{align}
    With this choice of the input distribution, we achieve the rate tuple
    \begin{align}\label{def: causal helper example tuple}
        (\frac{1}{2}H_b(\hat{\alpha}),H(D)-(1-\hat{\alpha}) H_b(\min\{\frac{D}{1-\hat{\alpha}},\frac{1}{2}\}),1-H_b(D)).
    \end{align}
    Then, it remains to be shown that a convex combination of these two tuples exists that is not achievable. To this end, we set $D=0.25$ and $\Delta_1 = 0.4$. The input distribution that achieves the tuple \eqref{def: causal helper example tuple} and the corresponding rates are 
    \begin{align}
        &\hat{\alpha} = 1-\frac{H_b(D)}{H_b(\Delta_1)}\approx 0.1644,\;\;\; H(\hat{\alpha}) \approx 0.6448,\\
        &H(D)-(1-\hat{\alpha}) H_b(\min\{\frac{D}{1-\hat{\alpha}},\frac{1}{2}\}) \approx  0.0757.
    \end{align}
    Now we consider the convex combination of the above two tuples with convex coefficients $(\frac{1}{2},\frac{1}{2})$. Assume the convex combination is also achievable. This implies that there exists an input distribution $\hat{\alpha}^*$ achieving
    \begin{align}
        (\frac{1}{2}H_b(\hat{\alpha}^*),H(D)-(1-\hat{\alpha}^*) H_b(\min\{\frac{D}{1-\hat{\alpha}^*},\frac{1}{2}\}),1-H_b(D))
    \end{align}
    such that
    \begin{align}
        &H_b(\hat{\alpha}^*)=\frac{1}{2}H_b(\hat{\alpha}),\\
        \label{eq: causal helper numerical example r1 alpha *}&H(D)-(1-\hat{\alpha}^*) H_b(\min\{\frac{D}{1-\hat{\alpha}^*},\frac{1}{2}\})= \frac{1}{2}\left(H(D)-(1-\hat{\alpha}) H_b(\min\{\frac{D}{1-\hat{\alpha}},\frac{1}{2}\})\right).
    \end{align}
    It follows that $\hat{\alpha}^*\in[0.05,0.06]$ and $D/(1-\hat{\alpha}^*)<\frac{1}{2}$ always holds. By the fact that $xH_b(\frac{1}{x})$ is a monotone increasing function of $x$, we substitute 0.05 and 0.06 into \eqref{eq: causal helper numerical example r1 alpha *} and the rate of the state helper in this case is
    \begin{align}
        R_{f_1} \in [0.0214,0.0258] \neq \frac{1}{2} \left( H(D)-(1-\hat{\alpha}) H_b(\min\{\frac{D}{1-\hat{\alpha}},\frac{1}{2}\}) \right).
    \end{align}
    Hence, this convex combination tuple is not achievable, and the region without time-sharing is generally not convex.
\end{remark}
For the converse, it follows that
\begin{align}
    R &\leq I(X;Y|S) = H(Y|S) - H(Y|XS) = P_S(1)H_b(X)\overset{(a)}{=}\frac{1}{2}H_b(\alpha),\\
    R_{f_1} &\geq I(V;S|X,Y,T) = H(S|X,Y,T) - H(S|X,Y,T,V)\\
    &=P_X(0)H(S|T,X=0) - H_b(S|X,Y,T,V,\hat{S})\\
    &\overset{(b)}{\geq} P_X(0)H(S|T,X=0) - (1-\alpha)H_b(S\oplus \hat{S}|X=0)\\
    &\overset{(c)}{=}(1-\alpha)H_b(\Delta_1)-(1-\alpha)H_b(\frac{D}{1-\alpha}),
\end{align}
where $(a)$ follows by setting $\alpha:=P_{X}(1)$, $(b)$ follows by noting that when $X=1$ we can always construct $\hat{S}=Y$ such that $\mathbb{E}[d(S,\hat{S})]=0$,  $(c)$ follows by setting $H_b(\Delta_1):=H(S|T,X=0)$ for some $0\leq \Delta_1 \leq \frac{1}{2}$.
\begin{align}
    R_{f_2} &\geq H(Y|X,V,T)\\
    &\overset{(a)}{=}P_X(1)H(S|V,T,X=1)=\alpha H_b(\Delta_2),\\
    R_{f_1} + R_{f_2} &\geq I(T;S|X) + I(V;S|X,Y,T) + H(Y|T,X)\\
    &=H(S) - H(S|X,T) + H(S|X,Y,T) - H(S|X,Y,T,V) + H(Y|T,X)\\
    &=H(S) - H(S|X,T) + P_X(0)H(S|T,X=0) -(1-\alpha) H_b(\frac{D}{1-\alpha}) + P_X(1)H(S|T,X=1)\\
    &=H(S)-(1-\alpha) H_b(\frac{D}{1-\alpha}), 
\end{align}
where $(a)$ follows by setting $H_b(\Delta_2) := H(S|V,T,X=1)$ for some $0\leq \Delta_2 \leq \frac{1}{2}$.

For the achievability part, we first consider the case that $R_{f_1}>0$ and $\frac{D}{1-\alpha}\leq \frac{1}{2}$. Note that this implies that $\Delta_1 > \frac{D}{1-\alpha}$. It will suffice to show the achievability of the following two sets of points:
\begin{itemize}
    \item $(\frac{1}{2}H_b(\alpha),1-(1-\alpha)H_b(\frac{D}{1-\alpha})-\alpha H_b(\Delta_2),\alpha H_b(\Delta_2))$;
    \item $(\frac{1}{2}H_b(\alpha),(1-\alpha)H_b(\Delta_1)-(1-\alpha)H_b(\frac{D}{1-\alpha}),1-(1-\alpha)H_b(\Delta_1))$;
\end{itemize}
which correspond to the following two corner points:
\begin{itemize}
    \item $(I(X;Y|S), I(T;S|X) + I(V;S|X,T),H(Y|X,V,T))$;
    \item $(I(X;Y|S), I(V;S|X,Y,T),I(T;S|X) + H(Y|X,T))$;
\end{itemize}
We construct the auxiliary random variables as follows. When $X=0$, we choose $V:= S \oplus N$, where $N\sim Bernoulli(P_N),P_N = \frac{D}{1-\alpha}$. Furthermore, we construct a binary $T$ such that $T-V-S$ forms a cascade channel and the transition probability from $T$ to $V$ is a binary symmetric channel with crossover probability $\beta=\frac{\Delta_1 - \frac{D}{1-\alpha}}{1-2\cdot \frac{D}{1-\alpha}}$. It can be easily verified that $\beta$ is a valid probability, and the crossover probability from $T$ to $S$ given $X=0$ is $\Delta_1$. When $X=1$, we simply construct a pair of $(V,T)$ such that $H_b(\Delta_2) = H(S|V,T,X=1)$ is satisfied. The state estimator is defined as
\begin{align}
    \hat{s} = \left\{
        \begin{aligned}
            y,\;\;\;  x=1,\\
            v,\;\;\;  x=0.
        \end{aligned}
    \right.
\end{align}

Now, we verify that the above corner points are achieved with these settings. The choice of $(V,T)$ is irrelevant for the bound on $I(X;Y|S)$ and hence $\frac{1}{2}H_b(\alpha)$ is always achieved once the input distribution is fixed. Then, it follows that
\begin{align}
    I(T;S|X) + I(V;S|X,T) &= I(V,T;S|X)\\
    &=H(S) - H(S|X,V,T)\\
    &=H(S) - H(S|X,V,T,Y) - I(S;Y|X,V,T)\\
    &= H(S) - (1-\alpha) H_b(\frac{D}{1-\alpha}) - H(Y|X,V,T) \\
    &=1 - (1-\alpha) H_b(\frac{D}{1-\alpha}) - \alpha H_b(\Delta_2)
\end{align}
and
\begin{align}
    H(Y|X,V,T) = P_X(1)H(S|V,T,X=1)=\alpha H_b(\Delta_2).
\end{align}
For the second set of points, we have
\begin{align}
    I(V;S|X,Y,T) &= H(S|X,Y,T) - H(S|X,Y,T,V)\\
    &= P_X(0)H(S|T,X=0) - (1-\alpha) H_b(\frac{D}{1-\alpha}) = (1-\alpha)H_b(\Delta_1) - (1-\alpha) H_b(\frac{D}{1-\alpha})
\end{align}
and
\begin{align}
    I(T;S|X) + H(Y|X,T) &= H(S) - H(S|X,T) + H(Y|X,T)\\
    &=H(S) - H(S|X,T) + P_X(1)H(S|T,X=1)\\
    &=H(S) - P_X(0)H(S|T,X=0) = 1 - (1-\alpha)H_b(\Delta_1).
\end{align}
The distortion is bounded by
\begin{align}
    \mathbb{E}[d(S,\hat{S})] &= \mathbb{E}[\mathbb{E}[d(S,\hat{S})]|X]\\
    &=P_X(0)\mathbb{E}[d(S,\hat{S})|X=0]\\
    &=D.
\end{align}

When $\frac{D}{1-\alpha}>\frac{1}{2}$, we set $V:= S \oplus N$, where $N\sim Bernoulli(P_N),P_N = \frac{1}{2}$. The distortion in this case is $\frac{1-\alpha}{2}<D$.

It remains to prove the case that the rate $R_{f_1}$ is 0. When
\begin{align}
    (1-\alpha)H_b(\Delta_1)-(1-\alpha) H_b(\min\{\frac{D}{1-\alpha},\frac{1}{2}\}) \leq 0,
\end{align}
we set $V=\emptyset$ and $T = S\oplus N, \hat{S}=T$, where $N\sim Bernoulli(P_N),P_N = \frac{D}{1-\alpha}$. Due to the inequality
\begin{align}
    1-(1-\alpha)H_b(\min\{\frac{D}{1-\alpha},\frac{1}{2}\})\geq \alpha \geq \alpha H_b(\Delta_2),
\end{align}
in this case, we only need to consider the sum rate constraint
\begin{align}
    1-(1-\alpha) H_b(\min\{\frac{D}{1-\alpha},\frac{1}{2}\}).
\end{align}
It follows that
\begin{align}
    I(S;T|X) + H(Y|X,T) &= I(S;T|X,Y) + H(Y|X)\\
    &= H(Y|X) + H(S|X,Y) - H(S|X,Y,T)\\
    &= \alpha H(S) + (1-\alpha) H(S) - (1-\alpha)H(S|X=0,T)\\
    &=1 - (1-\alpha)H(\frac{D}{1-\alpha}).
\end{align}
The case for $\frac{D}{1-\alpha} > \frac{1}{2}$ follows similarly and the proof is completed.

\section{proof of theorem \ref{the: inner bound}}\label{sec: proof of the general inner bound}
In the following, we describe the coding scheme for Theorem 1.

\subsection{Coding Scheme}
\emph{Codebook generation at the state helper:} the state helper generates $N_{11}=\exp\{n(I(S;V_1)+\epsilon)\}$ independent codewords $\{v^n_{11}\}$ for some $\epsilon>0$ indexed by $1\leq n_{11}\leq N_{11}$, each according to the distribution $P_{V_{11}}$. For each $v^n_{11}(n_{11})$, generate $\widetilde{N}_{12}=\exp\{n(\widetilde{R}_{f_{12}}+\epsilon)\}$ independent codewords $\{v_{12}^n|n_{11}\}$, each according to $P_{V_{12}|V_{11}}^n(v_{12}^n|v_{11}^n(n_{11}))=\prod_{i=1}^n P_{V_{12}|V_{11}}(v_{12,i}|v_{11,i}(n_{11}))$. Partition each codebook $\{v_{12}^n|n_{11}\}$ into $N_{12} = \exp\{n(R_{f_{12}}+\epsilon)\}$ bins $\{v_{12}^n|n_{11},n_{12}\}$ indexed by $(n_{11},n_{12}), 1\leq n_{12} \leq N_{12}$. Now each codeword $v^n_2\in\{v_{12}^n|n_{11}\}$ can be uniquely index by $(n_{12},l)$, where $l$ is the index of $v^n_{12}$ within the bin $(n_{11},n_{12})$. Denote the bin of a given $v_{12}^n$ by $b(v^n_{12})$.

\emph{Codebook generation at the sender: } The sender generates $\widetilde{M}=\exp\{n(I(U;Y)+\epsilon)\}$ independent codewords $\{u^n\}$, each according to $P_{U}$. It then partition the codebook into $M =\exp\{n(I(U;Y) - I(U;V_1)-\epsilon)\}$ bins $\{u^n|m\}$, each index by $1 \leq m \leq M$. Denote the bin of a given codeword $u^n$ by $b(u^n)$.

\emph{Codebook generation at the feedback helper: } the feedback helper generates $\widetilde{N}_2=\exp\{n(\widetilde{R}_{f_{2}}+\epsilon)\}$ independent codewords $\{v^n_2\}$, each according to the distribution $P_{V_2}$. It then partitions the codebook into $N_2=\exp\{n(R_{f_{2}}+\epsilon)\}$ bins $\{v^n_2|n_2\},1\leq n_2 \leq N_2$. Denote the bin number of a given codeword $v^n_2$ by $b(v^n_2)$.

\emph{Encoding at the state helper: } The indices sent by the state helper consist of two parts, which are denoted by $(\phi_{f_{11}}(s^n),\phi_{f_{12}}(s^n))$. At the beginning of each transmission block, the state helper observes the channel state $s^n$. It finds a codeword $v^n_{11}$ in $\{v^n_{11}\}$ with index $n_{11}$ such that $(s^n,v^n_{11})\in \mathcal{T}^n_{\delta}$. If there are multiple such codewords, the state helper selects the first one and then uses $n_{11}$ as the first index $\phi_{f_{11}}(s^n)$. If such a codeword does not exist, it sets $\phi_{f_{11}}(s^n)$ as the default number 1 and declares an error. After finding $v^n_{11}(n_{11})$, the state helper then finds a codeword $v^n_{12}\in\{v^n_{12}|n_{11}\}$ such that $(v^n_{11},v^n_{12},s^n)\in\mathcal{T}^n_{\delta}$. If there are multiple such codewords, the state helper selects the first one, and if such a codeword does not exist, the default number is selected, and the state helper declares an error. The second index $\phi_{f_{12}}(s^n)$ sent by the state helper is the bin $b(v^n_{12})$ of the selected codeword $v^n_{12}$. For simplicity, we also denote the selected codewords given $s^n$ by $v^n_{11}(s^n)$ and $v^n_{12}(s^n)$.

\emph{Encoding at Sender: } The sender receives two indices $(\phi_{f_{11}}(s^n),\phi_{f_{12}}(s^n))$ from the state helper at the beginning of the block. It finds $v^n_{11}$ according to the first index $\phi_{f_{11}}(s^n)$. To transmit the message $m$, it then looks for a codeword $u^n\in \{u^n|m\}$ such that $(u^n,v^n_{11})\in \mathcal{T}^n_{\delta}$. If there are multiple codewords that are jointly typical with $v^n_{11}$, the encoder selects the first one. Finally, the encoder generates  a codeword $x^n$ according to the distribution $P_{X|V_{11}U}^n(\cdot | v^n_{11}, u^n)$.

\emph{Encoding at the feedback helper:} At the end of the transmission block, the feedback helper receives all the feedback symbols $z^n$. It looks for a codeword $v^n_2$ such that $(z^n,v^n_2)\in\mathcal{T}^n_{\delta}$. If there are multiple codewords $v^n_2$ that are jointly typical with $z^n$, the feedback helper selects the first one. If there is no such codeword, the default codeword is chosen, and the feedback helper declares an error. After determining the codeword, the feedback helper sets its index $\phi_{f_2}(z^n)$ as the bin index to which $v^n_2$ belongs. We denote the selected codeword given $z^n$ by $v^n_2(z^n)$ for simplicity.

\emph{Decoding at the sender: } Although the sender receives two indices from the state helper at the beginning of the transmission block, it can only use the first one $\phi_{f_{11}}(s^n)$ to locate the codeword $\hat{v}^n_{11}$. It is not able to decode $\phi_{f_{12}}(s^n)$ due to the lack of side information until it receives $\phi_{f_2}(z^n)$ at the end of the current transmission block. It then looks for a unique pair of $(\hat{v}^n_{12},\hat{v}^n_2)$ in bins $\{v^n_{12}|\phi_{f_{11}}(s^n),\phi_{f_{12}}(s^n)\}\times \{v^n_2|\phi_{f_2}(z^n)\}$ such that $(\hat{v}^n_{11},\hat{v}^n_{12},\hat{v}^n_2,u^n)\in\mathcal{T}^n_{\delta}$.

\emph{Decoding at the receiver: } The receiver observes the channel output $y^n$. It looks for a unique $\hat{u}^n$ in the communication codebook such that $(u^n,y^n)\in\mathcal{T}^n_{\delta}$. If such a codeword does not exist or there are multiple such codewords, the receiver declares an error. Once the codeword $\hat{u}^n$ is found, the receiver declares that message $\hat{m}=b(\hat{u}^n)$ has been sent.

\emph{State Estimation: } The sender decodes codewords $(v^n_{11},v^n_{12},v^n_{2})$ using the indices $(\phi_{f_{11}}(s^n),\phi_{f_{12}}(s^n),\phi_{f_2}(z^n))$ sent by the state and feedback helpers. It estimates the state sequence using $\hat{s}(x,u,v_{12},v_2)$ elementwise.

\textbf{Error Analysis: }

In the following, we analyze the error probability in the coding scheme caused by encoding and decoding errors. Let $M$ be the sent message and $L_0$ be the index of $U^n$ within the bin $\mathcal{C}(M)$, and $B_{11},B_{12},B_{2}$ be the bin indices of codewords $(V^n_{11},V^n_{12},V^n_2)$ and $L_{11},L_{12},L_{2}$ be the indices of $(V^n_{11},V^n_{12},V^n_2)$ within the bins. We define error events as follows.

\emph{Encoding error at the state helper: }
\begin{align*}
    &\mathcal{E}_{11} =\left\{ \bigcap_{n_{11}=1}^{N_{11}}(V^n_{11}(n_{11}),S^n)\notin \mathcal{T}^n_{\delta}  \right\},\\
    &\mathcal{E}_{12} = \left\{ \bigcap_{\widetilde{n}_{12}=1}^{\widetilde{N}_{12}}(V^n_{11}(S^n),V^n_{12}(\widetilde{n}_{12}),S^n)\notin \mathcal{T}^n_{\delta}  \right\}.
\end{align*}
By the covering lemma \cite{el2011network}, we have $Pr\{\mathcal{E}_{11}\}\to 0$ and $Pr\{\mathcal{E}_{12}\}\to 0$ as $n\to \infty$ if 
\begin{align*}
    R_{f_{11}} > I(V_{11};S),\\
    \widetilde{R}_{f_{12}} > I(V_{12};S|V_{11}),
\end{align*}
and hence $Pr\{\mathcal{E}_1\}\to 0$, where $\mathcal{E}_1 := \mathcal{E}_{11} \cup \mathcal{E}_{12}$.

\emph{Encoding error at the sender: }
\begin{align*}
    &\mathcal{E}_0 = \left\{ \bigcap_{l_0=1}^{\widetilde{M}/M}(U^n(M,l_0),V^n_{11})\notin\mathcal{T}^n_{\delta} \right\}.
\end{align*}
Again by the covering lemma, we have $Pr\{\mathcal{E}_0\}\to 0$ as $n\to \infty$ by the setting of $\widetilde{M}$ and $M$.

\emph{Encoding error at the feedback helper: }
\begin{align*}
    &\mathcal{E}_2 = \left\{ \bigcap_{\widetilde{n}_{2}=1}^{\widetilde{N}_{2}}(V^n_{2}(\widetilde{n}_{2}),Z^n)\notin \mathcal{T}^n_{\delta} \right\}.
\end{align*}
Similarly, we have $Pr\{\mathcal{E}_2\} \to 0$ as $n\to \infty$ by setting
\begin{align*}
    \widetilde{R}_{f_2} > I(V_2;Z).
\end{align*}
\emph{Decoding error at the sender: } Next, we define the decoding error events at the sender side.
\begin{align*}
    &\mathcal{E}_3 = \left\{ (V^n_{11}(S^n),V^n_{12}(S^n),V^n_2(Z^n),U^n(M,L_0),S^n,Z^n) \notin \mathcal{T}^n_{\delta}   \right\},\\
    &\mathcal{E}_4 = \left\{ (V^n_{11}(B_{11},L_{11}),V^n_{12}(B_{12},L_{12}'),V^n_2(B_{2},L_{2}'),U^n(M,L_0)) \in \mathcal{T}^n_{\delta} \;\text{for some $(L_{12}',L_{2}')\neq (L_{12},L_{2})$}   \right\}.
\end{align*}
Note that $V_{11}-U-S$ and $U-(V_{11},S)-V_{12}$ form two Markov chains. By the covering lemma and by the setting of $R_{f_{11}}$ and $\widetilde{R}_{f_{12}}$, we have $Pr\{(V^n_{11}(S^n),V^n_{12}(S^n),S^n)\}\in\mathcal{T}^n_{\delta}\to 1$ as $n\to\infty$. Following the same argument as \cite[Lemma 12.3]{el2011network} we also have
\begin{align*}
    Pr\{U^n(M,L_0)=u^n|V^n_{11}(S^n)=v^n_{11}\}\overset{\sim}{=} 2^{nH(U|V_{11})},\\
    Pr\{V_{12}^n(S^n)=v_{12}^n|V^n_{11}(S^n)=v^n_{11},S^n=s^n\}\overset{\sim}{=} 2^{nH(V_{12}|V_{11}S)},
\end{align*}
which satisfies the second condition of the Markov Lemma. Hence, it follows that 
\begin{align*}
    Pr\{(S^n,V^n_{11}(S^n),V^n_{12}(S^n),U^n)\in\mathcal{T}^n_{\delta}\}\to 1\;\text{as $n\to\infty$.}
\end{align*}
As for the feedback signal $Z^n$, the distribution satisfies
\begin{align*}
    &Pr\{Z^n=z^n|V_{12}^n(S^n)=v_{12}^n,V^n_{11}(S^n)=v^n_{11},S^n=s^n,U^n(M,L_0)=u^n\}\\
    &= \sum_{x^n\in\mathcal{X}^n}P^n_{Z|XS}(z^n|x^n,s^n)P^n_{X|V_{11}U}(x^n|v^n_{11},u^n)\\
    &=\prod_{i=1}^n \sum_{x_i\in\mathcal{X}} P_{Z|XS}(z_i|x_i,s_i)P_{X|V_{11}U}(x_i|v_{11,i},u_i).
\end{align*}
By the above equality and the conditional typicality lemma\cite{el2011network}, we have
\begin{align*}
    Pr\left\{\bigcap_{i=0}^2 \mathcal{E}_i^c \cap \mathcal{E}_3\right\} \to 0
\end{align*}
as $n\to\infty$.
To bound $\mathcal{E}_4$, we invoke Lemma 10.6.2 in \cite{thomas2006elements} and consider the following three cases.
\begin{itemize}
    \item Only $L_{12}'$ is wrong: This case is upper bounded by
    \begin{align*}
        &Pr\left\{ (v^n_{11},V^n_{12}(B_{12},L_{12}'),v^n_2,u^n) \in \mathcal{T}^n_{\delta} \;\text{for some $L_{12}'\neq L_{12}$}   \right\}\\
        &\leq 2^{n(\widetilde{R}_{f_{12}}-R_{f_{12}})} \sum_{v^n_{12}\in\mathcal{T}^n_{\delta}[v^n_{11},u^n,v^n_{2}]}2^{-n(H(V_{12}|V_{11})+\epsilon(\delta))}\\
        & \leq 2^{n(\widetilde{R}_{f_{12}}-R_{f_{12}})}  2^{-n(I(V_{12};U,V_2|V_{11})-\epsilon(\delta))}.
    \end{align*}
    \item Only $L_{2}'$ is wrong: Similarly, this case is upper bounded by
    \begin{align*}
        &Pr\left\{ (v^n_{11},v^n_{12},V^n_2(B_2,L_2'),u^n) \in \mathcal{T}^n_{\delta} \;\text{for some $L_{12}'\neq L_{12}$}   \right\}\\
        &  \leq  2^{n(\widetilde{R}_{f_2}-R_{f_2})}\sum_{v^n_{2}\in\mathcal{T}^n_{\delta}[v^n_{11},u^n,v^n_{12}]}2^{-n(H(V_{2})-\epsilon(\delta))}\\
        &\leq 2^{n(\widetilde{R}_{f_2}-R_{f_2})} 2^{-n(I(V_{2};U,V_{11},V_{12})-\epsilon(\delta))}.
    \end{align*}
    \item Both $L_{12}'$ and $L_{2}'$ are wrong: 
    \begin{align*}
        &Pr\left\{ (v^n_{11},V^n_{12}(B_{12},L_{12}'),V^n_2(B_2,L_2'),u^n) \in \mathcal{T}^n_{\delta} \;\text{for some $L_{12}'\neq L_{12}, L_{2}'\neq L_{2}$}   \right\}\\
        &\leq 2^{n(\widetilde{R}_{f_{12}}-R_{f_{12}}+\widetilde{R}_{f_{12}}-R_{f_{12}})} \sum_{(v^n_{12},v^n_{2})\in\mathcal{T}^n_{\delta}[v^n_{11},u^n]}2^{-n(H(V_{2})+H(V_{12|V_{11}})-\epsilon(\delta))}\\
        &\leq 2^{n(\widetilde{R}_{f_{12}}-R_{f_{12}}+\widetilde{R}_{f_{2}}-R_{f_{2}})}2^{-n(H(V_{2})+H(V_{12}|V_{11})- H(V_{12},V_2|U,V_{11}) - \epsilon(\delta))}.
    \end{align*}
\end{itemize}
Thus, to ensure that we have
\begin{align*}
    Pr\left\{\bigcap_{i=0}^3 \mathcal{E}_i^c \cap \mathcal{E}_4\right\} \to 0
\end{align*}
the following inequalities should be satisfied:
\begin{align}
    \widetilde{R}_{f_{12}}-R_{f_{12}} &\leq I(V_{12};U,V_2|V_{11}),\\
    \widetilde{R}_{f_2}-R_{f_2} &\leq I(V_{2};U,V_{11},V_{12}),\\
    \widetilde{R}_{f_{12}}-R_{f_{12}}+\widetilde{R}_{f_{12}}-R_{f_{12}} &\leq H(V_{2})+H(V_{12}|V_{11})- H(V_{12},V_2|U,V_{11})\\
    &=H(V_2) + H(V_{12}|V_{11}) - H(V_{12}|U,V_{11}) - H(V_2|U,V_{11},V_{12})\\
    &\overset{(a)}{=}I(V_2;U,V_{11},V_{12}),
\end{align}
where $(a)$ follows by the Markov chain $V_{12}-V_{11}-U$.

\emph{Decoding error at the receiver: }
\begin{align*}
    &\mathcal{E}_5 = \left\{ (U^n(M,L_0),Y^n)\notin T^n_{\delta}  \right\},\\
    &\mathcal{E}_6 = \left\{ (U^n(M',L_0'),Y^n)\in T^n_{\delta}\;\text{for some $M'\neq M$}.  \right\}.
\end{align*}
The analysis of events $\mathcal{E}_5$ and $\mathcal{E}_6$ is the same as for the Gel'fand-Pinsker coding\cite{gel1980coding} and is arbitrarily small with $n\to \infty$ by conditional typicality lemma and setting
\begin{align*}
    &\widetilde{R} \leq I(U;Y).
\end{align*}

Now, using the Fourier-Motzkin Elimination gives
\begin{equation}\label{def: achievable region 2}
    \left\{
        \begin{aligned}
            & R \leq I(U;Y)-I(U;V_{11}),\\
            & R_{f_{11}} \geq I(V_{11};S),\\
            & R_{f_{12}} \geq I(V_{12};S,Z|U,V_{11},V_2),\\
            & R_{f_2} \geq I(V_2;S,Z|U,V_{11},V_{12}),\\
            & R_{f_{12}} + R_{f_2} \geq I(V_{12},V_2;S,Z|U,V_{11}),\\
            & R_{f_1} = R_{f_{11}} + R_{f_{12}}.
        \end{aligned}
    \right.
\end{equation}

\textbf{State Estimation:}

The distortion at the state estimator side can be bounded as
\begin{align*}
    \frac{1}{n}\mathbb{E}\left[d^n(S^n,\hat{S}^n)\right]\leq P_e\cdot d_{max} + (1+\epsilon)\mathbb{E}\left[d(S,\hat{S})\right] \leq D+ \epsilon',
\end{align*}
where the inequality follows by considering the distortion upper bound with or without error separately and the Typical Average Lemma\cite{el2011network}. The achievability proof is completed by the fact that $\epsilon'\to 0$ as $n\to\infty$.

\subsection{Cardinality bounds}
We consider the joint distribution $P_{SV_{11}V_{12}UXYZV_2|q}$ on $\mathcal{S}\times\mathcal{V}_{11}\times\mathcal{V}_{12}\times\mathcal{U}\times\mathcal{X}\times\mathcal{Y}\times\mathcal{Z}\times\mathcal{V}_{2}$ distributed as \eqref{def: joint distribution}. By the Fenchel-Eggleston-Carath\'eodory theorem we can preserve the value of $I(U;Y|Q)-I(U;V_{11}|Q), I(V_{11};S|Q)+ I(V_{12};S,Z|U,V_{11},V_2,Q),$ $I(V_2;S,Z|U,V_{11},V_{12},Q),I(V_{11};S|Q)+ I(V_{12},V_2;S,Z|U,V_{11},Q)$ and $\mathbb{E}\left[ d(S,\hat{S}(V_{11},V_{12},V_2,U,X,Q))  \right]$ by restricting the size of $|\mathcal{Q}|\leq 5$.

To bound the remaining alphabet sizes, we write $\mathcal{R}^{in}$ in \eqref{def: achievable region} as $\mathcal{R}^{in}_{|\mathcal{V}_{11}|}$ representing that the region is computed with auxiliary random variable $V_{11}$ and alphabet size $|\mathcal{V}_{11}|$. In the following, we show 
\begin{align}
    \label{ineq: equivalent inequality}\mathcal{R}^{in}_{|\mathcal{V}_{11}|} \subseteq \mathcal{R}^{in}_{|\mathcal{S}||\mathcal{X}|+1}.
\end{align}
It is sufficient to show $\bar{\mathcal{R}}^{in}_{|\mathcal{V}_{11}|} \subseteq \bar{\mathcal{R}}^{in}_{|\mathcal{S}||\mathcal{X}|+1}$, where $\bar{\mathcal{R}}^{in}_{|\mathcal{V}_{11}|}$ is the set of real tuples $(\bar{R}_1,\bar{R}_2,\bar{R}_3,\bar{R}_4,\bar{R}_5)$ such that 
\begin{align}
    &\bar{R}_1\leq I(U;Y|Q)-I(U;V_{11}|Q),\\
    &\bar{R}_2\geq I(V_{11};S|Q)+ I(V_{12};S,Z|U,V_{11},V_2,Q),\\
    &\bar{R}_3\geq I(V_2;S,Z|U,V_{11},V_{12},Q),\\
    &\bar{R}_4\geq I(V_{11};S|Q)+ I(V_{12},V_2;S,Z|U,V_{11},Q),\\
    &\bar{R}_5 \geq  \mathbb{E}\left[ d(S,\hat{S}(V_{12},V_2,U,X,Q))  \right].
\end{align}
To this end, due to the introduction of the time-sharing random variable $Q$ and the fact that $\mathbb{E}\left[ d(S,\hat{S}(V_{11},V_{12},V_2,U,X,Q))  \right]$ is a linear function of the joint distribution \eqref{def: joint distribution}, it suffices to consider the following inequality with any real tuples $(\lambda_1,\lambda_2,\lambda_3,\lambda_4,\lambda_5)\in\mathbb{R}^5$:
\begin{align}
    \max_{(\bar{R}_1,\bar{R}_2,\bar{R}_3,\bar{R}_4,\bar{R}_5)\in \mathcal{R}^{in}_{|\mathcal{V}_{11}|}}\sum_{i=1}^5 \lambda_i \bar{R}_i \leq \max_{(\bar{R}_1,\bar{R}_2,\bar{R}_3,\bar{R}_4,\bar{R}_5)\in \mathcal{R}^{in}_{|\mathcal{S}||\mathcal{X}|+1}} \sum_{i=1}^5 \lambda_i \bar{R}_i.
\end{align}
Without loss of generality, we only consider non-negative $\lambda_1$. Otherwise, one can set $\widetilde{R}_1=-\infty$ such that both sides converge to $\infty$ and the equality holds. For the same reason, we consider only non-positive $\lambda_i,i=2,3,4,5$. 
Define
\begin{align}
    C(P)&=\lambda_1 (I(U;Y)-I(U;V_{11})) + \lambda_2(I(V_{11};S|Q)+ I(V_{12};S,Z|U,V_{11},V_2,Q)) + \lambda_3 (I(V_2;S,Z|U,V_{11},V_{12},Q))\\
    &+\lambda_4 (I(V_{11};S|Q)+ I(V_{12},V_2;S,Z|U,V_{11},Q))
\end{align}
under the joint distribution $P$ satisfying \eqref{def: joint distribution}.

Consider the perturbed distribution
\begin{align}
    P_{\epsilon}(s,v_{11},v_{12},u,x,y,z,v_{12},q) = P_{0}(s,v_{11},v_{12},u,x,y,z,v_{12},q) (1+L(v_{11})),
\end{align}
where $P_0:=\mathop{\arg\max}_{P} \left( C(P) + \lambda_5(\mathbb{E}\left[ d(S,\hat{S}(V_{12},V_2,U,X,Q))  \right]) \right)$. The function $L:\mathcal{V}_{11}\to\mathbb{R}$ satisfies
\begin{align}
    \label{def: l1}\mathbb{E}\left[L(V_{11}) \right]=0,\\
    \label{def: l2}\mathbb{E}\left[L(V_{11}|s,x) \right]=0,\;\;\text{for all $(s,x)\in\mathcal{S}\times\mathcal{X}.$}
\end{align}
where
\begin{align}
    \mathbb{E}\left[L(V_{11}) \right] = \sum_{v_{11}}P_0(v_{11})L(v_{11}),\\
    \mathbb{E}\left[L(V_{11}|S,X) \right] = \sum_{v_{11},x,s}P_0(s,x)P_0(v_{11}|s,x)L(v_{11}).
\end{align}
Such a function $L$ exists if $|\mathcal{V}_{11}|\geq |\mathcal{S}||\mathcal{X}|+1.$ The function \eqref{def: l2} implies the distribution of $(S,X)$ is also preserved, and hence also the joint distribution of $P_0(x,s,y,z)$ for $(x,s,y,z)\in\mathcal{X}\times\mathcal{S}\times\mathcal{Y}\times\mathcal{Z}$. 

Define a set of random variables $(\bar{S},\bar{V}_{11},\bar{V}_{12},\bar{U},\bar{X},\bar{Y},\bar{Z},\bar{V}_{2},\bar{Q})\sim P_{\epsilon}$. For simplicity, we omit $Q$ in the following argument. Then, by \cite[Lemma 2]{gohari2012evaluation},
\begin{align}
    &I(\bar{U};\bar{Y}) - I(\bar{U};\bar{V}_{11}) \\
    &= H(\bar{Y}) - H(\bar{Y},\bar{U})-H(\bar{V}_{11}) + H(\bar{V}_{11},\bar{U})\\
    &= H(Y) - H(\bar{Y},\bar{U})-H(V_{11})+ H(V_{11},U)+\epsilon (H_L(V_{11},U)-H_L(V_{11}) )\\
    &I(\bar{V}_{11};\bar{S})+ I(\bar{V}_{12};\bar{S},\bar{Z}|\bar{U},\bar{V}_{11},\bar{V}_2) \\
    &= H(\bar{S}) - H(\bar{S},\bar{V}_{11})+H(\bar{V}_{11}) + H(\bar{S},\bar{Z},\bar{U},\bar{V}_{11},\bar{V}_2) - H(\bar{U},\bar{V}_{11},\bar{V}_2)- H(\bar{S},\bar{Z},\bar{U},\bar{V}_{11},V_{12},\bar{V}_2) + H(\bar{U},\bar{V}_{11},\bar{V}_{12},\bar{V}_2)\\
    &=H(S)-H(S,V_{11})+H(V_{11})+H(S,Z,U,V_{11},V_2)-H(U,V_{11},V_2)-H(S,Z,U,V_{11},V_{12},V_2)+H(U,V_{11},V_{12},V_2)\\
    &\quad +\epsilon(H_L(V_{11})-H_L(S,V_{11})+H_L(S,Z,U,V_{11},V_2)-H_L(U,V_{11},V_2)-H_L(S,Z,U,V_{11},V_{12},V_2)+H_L(U,V_{11},V_{12},V_2))\\
    &I(\bar{V}_2;\bar{S},\bar{Z}|\bar{U},\bar{V}_{11},\bar{V}_{12})\\
    &=H(\bar{S},\bar{Z},\bar{U},\bar{V}_{11},\bar{V}_{12}) - H(\bar{U},\bar{V}_{11},\bar{V}_{12}) - H(\bar{S},\bar{Z},\bar{U},\bar{V}_{11},\bar{V}_{12},\bar{V}_2)+H(\bar{U},\bar{V}_{11},\bar{V}_{12},\bar{V}_2)\\
    &=H(S,Z,U,V_{11},V_{12}) - H(U,V_{11},V_{12})- H(S,Z,U,V_{11},V_{12},V_2) + H(U,V_{11},V_{12},V_2)\\
    &\quad\quad\quad\quad\quad + \epsilon(H_L(S,Z,U,V_{11},V_{12}) - H_L(U,V_{11},V_{12})- H_L(S,Z,U,V_{11},V_{12},V_2) + H_L(U,V_{11},V_{12},V_2))
\end{align}
\begin{align}
    &I(\bar{V}_{11};\bar{S})+ I(\bar{V}_{12},\bar{V}_2;\bar{S},\bar{Z}|\bar{U},\bar{V}_{11})\\
    &= H(\bar{S}) - H(\bar{S},\bar{V}_{11})+H(\bar{V}_{11}) + H(\bar{S},\bar{Z},\bar{U},\bar{V}_{11})-H(\bar{U},\bar{V}_{11}) - H(\bar{S},\bar{Z},\bar{U},\bar{V}_{11},\bar{V}_{12},\bar{V}_2) + H(\bar{U},\bar{V}_{11},\bar{V}_{12},\bar{V}_2)\\
    &=H(S)-H(S,V_{11})+H(V_{11})+H(S,Z,U,V_{11})-H(U,V_{11})-H(S,Z,U,V_{11},V_{12},V_2)+H(U,V_{11},V_{12},V_2)\\
    &\quad +\epsilon(H_L(V_{11})-H_L(S,V_{11})+H_L(S,Z,U,V_{11})-H_L(U,V_{11})-H_L(S,Z,U,V_{11},V_{12},V_2)+H_L(U,V_{11},V_{12},V_2))
\end{align}

Since $P_0$ maximizes function $C(P) + \lambda_5(\mathbb{E}\left[ d(S,\hat{S}(V_{12},V_2,U,X,Q))  \right])$, the first and second derivatives satisfy
\begin{align}
    \frac{\partial}{\partial \epsilon}\left( C(P_\epsilon) + \lambda_5(\mathbb{E}_{P_\epsilon}\left[ d(S,\hat{S}(V_{12},V_2,U,X,Q))  \right]) \right)\bigg|_{\epsilon=0}=0,\\
    \frac{\partial^2}{\partial \epsilon^2}\left( C(P_\epsilon) + \lambda_5(\mathbb{E}_{P_\epsilon}\left[ d(S,\hat{S}(V_{12},V_2,U,X,Q))  \right]) \right)\bigg|_{\epsilon=0}\leq 0.
\end{align}
The latter reduces to
\begin{align}
    -\lambda_1 \frac{\partial^2}{\partial \epsilon^2}H(\bar{U},\bar{Y})\bigg|_{\epsilon=0}\leq 0.
\end{align}
By \cite[Lemma 2, Part 2]{gohari2012evaluation} and the non-negativeness of $\lambda_1$, it implies that
\begin{align}
    \mathbb{E}\left[ \left( \mathbb{E}\left[ L(V_{11})|U,Y \right] \right)^2 \right] \leq 0
\end{align}
with equality holding if
\begin{align}
    \mathbb{E}\left[ L(V_{11})|U=u,Y=y \right]=0
\end{align}
for each $(u,y)$ such that $P_0(u,y)>0$. Thus, we have $P_{\epsilon}(u,y)=P_0(u,y)$ and $H(\bar{U},\bar{Y})=H(U,Y)$. We can now write the function $C(P_{\epsilon})$ as
\begin{align}
    &C(P_{\epsilon})+\lambda_5\cdot \mathbb{E}_{P_\epsilon}\left[ d(S,\hat{S}(V_{11},V_{12},V_2,U,X,Q))  \right]\notag \\
     &= C(P_0)+ \lambda_5\mathbb{E}_{P_0}\left[ d(S,\hat{S}(V_{11},V_{12},V_2,U,X,Q))  \right]  + \lambda_1\cdot\epsilon (H_L(V_{11},U)-H_L(V_{11}) )\notag\\
    &+\lambda_2 \cdot\epsilon(H_L(V_{11})-H_L(S,V_{11})+H_L(S,Z,U,V_{11},V_2)-H_L(U,V_{11},V_2)-H_L(S,Z,U,V_{11},V_{12},V_2)+H_L(U,V_{11},V_{12},V_2))\notag \\
    &+\lambda_3 \cdot\epsilon(H_L(S,Z,U,V_{11},V_{12}) - H_L(U,V_{11},V_{12})- H_L(S,Z,U,V_{11},V_{12},V_2) + H_L(U,V_{11},V_{12},V_2))\notag \\
    &+\lambda_4 \cdot\epsilon(H_L(V_{11})-H_L(S,V_{11})+H_L(S,Z,U,V_{11})-H_L(U,V_{11})-H_L(S,Z,U,V_{11},V_{12},V_2)+H_L(U,V_{11},V_{12},V_2))\notag \\
    &+\lambda_5\cdot \epsilon \sum_{s,v_{11},v_{12},u,x,v_{12}}\sum_{y,z}P_0(s,v_{11},v_{12},u,x,y,z,v_{12})L(v_{11})d(s,\hat{s}).\notag
\end{align}
The first derivative condition \eqref{def: l1} implies
\begin{align}
    C(P_{\epsilon})+\lambda_5\cdot \mathbb{E}_{P_\epsilon}\left[ d(S,\hat{S}(V_{12},V_2,U,X,Q))  \right] = C(P_0)+ \lambda_5\mathbb{E}_{P_0}\left[ d(S,\hat{S}(V_{12},V_2,U,X,Q))  \right].
\end{align}
Hence, restricting the alphabet size of $\mathcal{V}_{11}$ to $|\mathcal{S}||\mathcal{X}|+1$ does not reduce our achievable region $\mathcal{R}^{in}(D)$. Similarly, we bound the alphabet size of $\mathcal{U}$ as $|\mathcal{U}|=|\mathcal{S}||\mathcal{X}||\mathcal{V}_{11}|+1=|\mathcal{S}||\mathcal{X}|(|\mathcal{S}||\mathcal{X}|+1)+1$( we should take into account the size of $\mathcal{V}_{11}$ as well. Otherwise, the value of $I(V_{11};S)$ cannot be preserved). Once the sizes of $\mathcal{U}$ and $\mathcal{V}_{11}$ are determined, we can bound the size of $\mathcal{V}_{12}$ and $\mathcal{V}_2$ using standard technique based on the Fenchel-Eggleston-Carath\'eodory theorem, which gives $\mathcal{V}_{12}\leq |\mathcal{S}||\mathcal{X}||\mathcal{V}_{11}|+4=|\mathcal{S}||\mathcal{X}|(|\mathcal{S}||\mathcal{X}|+1)+4$ and $\mathcal{V}_2 \leq |\mathcal{Z}|+4$.
\section{converse proofs}\label{sec: converse proofs}

In this section, we provide the converse proofs of the previous results.
\subsection{Converse of Corollary \ref{coro: rate unlimited feedback}}\label{sec: converse of rate unlimited feedback}

In this section, we use $M_{f_1}$ as the message sent by the state helper, $f_1$ as the state helper's encoder, and $f$ as the message sender's encoder.

To prove the converse bound in Corollary \ref{coro: rate unlimited feedback}, assume there exists a code $(n,R,R_{f_1})$ such that
\begin{align}
    &P_e \to 0,\\
    \label{inq: converse distortion constraint}&\mathbb{E}\left[ d(S,h(f_1(S^n),X^n,Y^n)) \right]\leq D.
\end{align}

The following lemma on the Markov chain will be useful in the proof of the converse.
\begin{lemma}{\cite[Proposition 2.5]{yeung2008information}}
    Two random variables $X$ and $Y$ are conditionally independent given $Z$ if and only if the joint distribution of $(X,Y,Z)$ can be written as 
    \begin{align*}
        P_{XYZ}(x,y,z) = g_1(x,z)\cdot g_2(y,z).
    \end{align*}
\end{lemma}
For any code $(n,R,R_{f_1})$, the joint distribution satisfies
\begin{align*}
    P(s^n,m_f,m,x^n,y^n) = P_S^n(s^n)\mathbb{I}\{m_f = f_1(S^n)\}P_M(m) P(x^n|m,m_f) \prod_{i=1}^n P_{Y|XS}(y_i|x_i,s_i).
\end{align*}
To bound $R_{f_1}$ we have
\begin{align}
    nR_{f_1} &\geq H(M_{f_1})\notag \\
    &\geq I(M_{f_1};S^n)\notag \\
    &=I(M_{f_1},X^n,Y^n;S^n) - I(X^n,Y^n;S^n|M_{f_1})\notag \\
    &=\sum_{i=1}^n I(M_{f_1},X^n,Y^n;S_i|S^{i-1}) - I(X_i,Y_i;S^n|X^{i-1},Y^{i-1},M_{f_1})\notag \\
    \label{eq: converse 1}&\overset{(a)}{=}\sum_{i=1}^n I(M_{f_1},X^n,Y^n,S^{i-1};S_i)  - I(X_i,Y_i;S^n|X^{i-1},Y^{i-1},M_{f_1}),
\end{align}
where $(a)$ follows by the i.i.d. property of the channel state. Now, we bound the second term in \eqref{eq: converse 1}. It follows that
\begin{align}
    &I(X_i,Y_i;S^n|X^{i-1},Y^{i-1},M_{f_1}) \\
    &=I(X_i;S^n|X^{i-1},Y^{i-1},M_{f_1}) + I(Y_i;S^n|X^{i},Y^{i-1},M_{f_1}) \\
    \label{eq: converse 5}&\overset{(a)}{=} H(Y_i|X^{i},Y^{i-1},M_{f_1}) - H(Y_i|X^{i},Y^{i-1},M_{f_1},S^n) \\
    &\overset{(b)}{\leq} H(Y_i|X_{i},M_{f_1}) - H(Y_i|X_{i},M_{f_1},S_i) \\
    \label{eq: converse 2}&= I(Y_i;S_i|X_{i},M_{f_1}).
\end{align}
where $(a)$ follows by the Markov chain $X_i-(X^{i-1},Y^{i-1},M_{f_1})-S^n$. The Markov chain relation can be shown as
\begin{align*}
    &P(s^n,x_i,x^{i-1},y^{i-1},m_{f_1})\\
    &=P_S^{n}(s^{n})\mathbb{I}\{m_f = f_1(s^n)\}\sum_{m}P_M(m) P(x^i|m,m_{f_1}) \prod_{k=1}^{i-1} P_{Y|XS}(y_k|x_k,s_k)\\
    &=\left( \sum_{m}P_M(m)P(x^i|m,m_{f_1}) \right)\left( P_S^{n}(s^{n})\mathbb{I}\{m_f = f_1(s^n)\} \prod_{k=1}^{i-1} P_{Y|XS}(y_k|x_k,s_k)\right),
\end{align*}
$(b)$ follows by the fact that condition reduces entropy and the Markov chain $Y_i-(X_i,S_i)-(X^{i-1},Y^{i-1},M_{f_1})$. 

Now we proceed to bound \eqref{eq: converse 1}. Substituting \eqref{eq: converse 2} into \eqref{eq: converse 1} gives
\begin{align}
    nR_{f_1}&\geq \sum_{i=1}^nI(M_{f_1},X^n,Y^n,S^{i-1};S_i) - I(Y_i;S_i|X_{i},M_{f_1})\\
    &=\sum_{i=1}^n I(M_{f_1},X_i;S_i) + I(S^{i-1},X^{n\backslash i},Y^n;S_i|X_{i},M_{f_1}) - I(Y_i;S_i|X_{i},M_{f_1})\\
    &=\sum_{i=1}^n  I(M_{f_1},X_i;S_i) + I(S_i;S^{i-1},X^{n\backslash i},Y^{n\backslash i}|X_i,Y_i,M_{f_1})\\
    &\geq \sum_{i=1}^n  I(M_{f_1};S_i) + I(S_i;S^{i-1},X^{n\backslash i},Y^{n\backslash i}|X_i,Y_i,M_{f_1}) \\
    \label{ine: sum rate rf1}&\overset{(a)}{=}\sum_{i=1}^nI(V_{11,i};S_i) + I(S_i;V_{12,i}|X_i,Y_i,V_{11,i})\\
    &=n\cdot \frac{1}{n}\sum_{i=1}^n(I(V_{11,Q};S_Q|Q=i) + I(S_Q;V_{12,Q}|X_Q,Y_Q,V_{11,Q},Q=i))\\
    &=n(I(V_{11,Q};S_Q|Q) + I(S_Q;V_{12,Q}|X_Q,Y_Q,V_{11,Q},Q))\\
    &\overset{(b)}{=}n(I(V_{11,Q},Q;S_Q) + I(S_Q;V_{12,Q}|X_Q,Y_Q,V_{11,Q},Q))\\
    \label{eq: converse 4}&\overset{(c)}{=}n(I(V_{11};S) + I(S;V_{12}|X,Y,V_{11}))
\end{align}
where $(a)$ follows by setting $V_{11,i}=M_{f_1},V_{12,i}=(S^{i-1},X^{n\backslash i},Y^{n\backslash i},V_{11,i})$, $(b)$ follows by the i.i.d. property of the channel state, $(c)$ follows $V_{11}=(V_{11,Q},Q),V_{12}=V_{12,Q}$.
On the other hand, we have
\begin{align*}
    nR &\leq H(M)\\
    &\leq I(M;Y^n|S^n) + n\delta\\
    &=I(M;Y^n|S^n,M_{f_1}) + n\delta\\
    &=\sum_{i=1}^n I(M;Y_i|Y^{i-1},S_i,S^{n\backslash i},M_{f_1})+ n\delta\\
    &\leq \sum_{i=1}^n I(S^{n\backslash i},Y^{i-1},M;Y_i|S_i,M_{f_1})+ n\delta\\
    &\leq \sum_{i=1}^{n} I(X_i;Y_i|S_i,M_{f_1})+ n\delta\\
    &=\sum_{i=1}^{n} I(X_i;Y_i|S_i,V_{11,i})+ n\delta\\
    &=nI(X;Y|S,V_{11}) + n\delta,
\end{align*}
where $V_{11,i}=M_{f_1}$.

\subsection{Converse of Theorem \ref{the: common component inner and outer bounds}}\label{sec: converse common component}
For lossless reconstruction of the feedback signal, the bound of $R_{f_1}$ is a little different from \eqref{eq: converse 4}. We start from \eqref{eq: converse 5} and it follows that
\begin{align}
    &I(X_i,Y_i;S^n|X^{i-1},Y^{i-1},M_{f_1}) \\
    &=H(Y_i|X^{i},Y^{i-1},M_{f_1}) - H(Y_i|X^{i},Y^{i-1},M_{f_1},S^n)\\
    &\overset{(a)}{=}H(Y_i|X^{i},Y^{i-1},M_{f_1}) - H(Y_i|X^{i},Y^{i-1},M_{f_1},S_i)\\
    &=I(Y_i;S_i|X^{i},Y^{i-1},M_{f_1}),
\end{align}
where $(a)$ is by the Markov chain $Y_i-(X^{i},Y^{i-1},M_{f_1},S_i)-S^{n\backslash i}$,
and
\begin{align}
    nR_{f_1}\geq H(M_{f_1}) &\geq  \sum_{i=1}^nI(M_{f_1},X^n,Y^n,S^{i-1};S_i) - I(Y_i;S_i|X^{i},Y^{i-1},M_{f_1})\\
    &\geq \sum_{i=1}^n  I(M_{f_1},X^i,Y^{i-1};S_i) + I(S^{i-1},X^{n}_{i+1},Y^n_{i};S_i|X^i,Y^{i-1},M_{f_1})-I(Y_i;S_i|X^{i},Y^{i-1},M_{f_1})\\
    \label{eq: converse}&=\sum_{i=1}^n  I(M_{f_1},X^i,Y^{i-1};S_i) + I(S^{i-1},X^{n}_{i+1},Y^n_{i+1};S_i|X^i,Y_i,Y^{i-1},M_{f_1})\\
    \label{eq: converse 7}&\overset{(a)}{=}\sum_{i=1}^n  I(M_{f_1},X^i,Y^{i-1};S_i) + I(S^{i-1},Y^n_{i+1};S_i|X^i,X^{n}_{i+1},Y_i,Y^{i-1},M_{f_1})\\
    &\geq \sum_{i=1}^n  I(M_{f_1};S_i) + I(S^{i-1},Y^n_{i+1};S_i|X^i,X^{n}_{i+1},Y_i,Y^{i-1},M_{f_1})
\end{align}
where $(a)$ follows by the Markov chain $S_i-(X^i,Y^i,M_{f_1})-X^n_{i+1}$.
Now, we define $V_{11,i}=M_{f_1},T_{i}=(X^{n\backslash i},Y^{i-1})$ and redefine $V_{12}$ as $(S^{i-1},Y_{i+1}^n,V_{11,i},T_i)$. To bound $M_{f_2}$, it follows that 

\begin{align}
    nR_{f_2} &\geq H(M_{f_2})\\
    &\geq H(M_{f_2}|M_{f_1},X^n) + H(Y^n|M_{f_1},M_{f_2},X^n) - H(Y^n|M_{f_1},M_{f_2},X^n)\\
    &\overset{(a)}{\geq} H(Y^n|X^n,M_{f_1}) - n\epsilon\\
    \label{eq: converse 3}&=\sum_{i=1}^n H(Y_i|Y^{i-1},X^n,M_{f_1}) - n\epsilon \\
    &\geq \sum_{i=1}^n H(Y_i|Y^{n \backslash i},X^n,M_{f_1},S^{i-1}) - n\epsilon \\
    &=\sum_{i=1}^n H(Y_i|V_{11,i},V_{12,i},X_i,T_i) - n\epsilon \\
    \label{eq: converse 6}&=n H(Y|V_{11},V_{12},X,T) - n\epsilon,
\end{align}
where $(a)$ is by the Fano's inequality and $H(Y^n|M_{f_1},M_{f_2},X^n) \leq n\epsilon$, the last inequality follows by introducing a time-sharing random variable $Q$.
To bound the sum rate
\begin{align}
    &n(R_{f_1}+R_{f_2}) \\
    &\geq H(M_{f_1},M_{f_2})\\
    &=H(M_{f_1}) + H(M_{f_2}|M_{f_1}) \\
    &\overset{(a)}{\geq}\sum_{i=1}^n I(M_{f_1},X^i,Y^{i-1};S_i) + I(S_i;S^{i-1},X^{n}_{i+1},Y^n_{i+1}|X^i,Y_i,Y^{i-1},M_{f_1}) + H(M_{f_2}|M_{f_1})\\
    &\overset{(b)}{=}\sum_{i=1}^nI(M_{f_1},X^{n},Y^{i-1};S_i) + I(S_i;S^{i-1},Y^n_{i+1}|X^{n}_{i+1},X^i,Y_i,Y^{i-1},M_{f_1}) + H(M_{f_2}|M_{f_1})\\
    &=\sum_{i=1}^nI(M_{f_1},X_i;S_i) + I(X^{n \backslash i},Y^{i-1};S_i|M_{f_1},X_i) + I(S_i;S^{i-1},Y^n_{i+1}|X^{n}_{i+1},X^i,Y_i,Y^{i-1},M_{f_1}) + H(M_{f_2}|M_{f_1})\\
    &\overset{(c)}{\geq}\sum_{i=1}^n I(M_{f_1};S_i) + I(X^{n \backslash i},Y^{i-1};S_i|M_{f_1},X_i) + I(S_i;S^{i-1},Y^n_{i+1}|X^{n}_{i+1},X^i,Y_i,Y^{i-1},M_{f_1}) + H(Y_i|Y^{i-1},X^n,M_{f_1}) - n\epsilon\\
    &=\sum_{i=1}^n I(V_{11,i};S_i) +I(T_i;S_i|V_{11,i},X_i) + I(V_{12,i};S_i|T_i,X_i,Y_i,V_{11,i}) + H(Y|T_i,X_i,V_{11,i})- n\epsilon\\
    &=n(I(V_{11};S)+I(T;S|V_{11},X)+I(V_{12};S|T,X,Y,V_{11}) + H(Y|T,V_{11},X)-\epsilon),
\end{align}
where $(a)$ follows from \eqref{eq: converse}, $(b)$ follows by \eqref{eq: converse 7} and the Markov chain $S_i-(M_{f_1},X^{i-1},Y^{i-1})-X_{i}^n$, which is proved by
\begin{align}
    &P(m_{f_1},s_{i},y^{i-1},x^n)\\
    &= \sum_{s^{n\backslash i}}P(s^n)P(m_{f_1}|s^n)\sum_{m}P(m)P(x^n|m,m_{f_1}) \prod_{j=1}^{i-1} P(y_j|x_j,s_j)\\
    &=\left(\sum_{s^{n\backslash i}}P(s^n)P(m_{f_1}|s^n) \prod_{j=1}^{i-1} P(y_j|x_j,s_j)\right) \sum_{m}P(m)P(x^n|m,m_{f_1}),
\end{align}
and the last inequality $(c)$ is due to \eqref{eq: converse 6}.

For the communication rate, it follows that
\begin{align*}
    nR &\leq H(M)\\
    &\leq I(M;Y^n|S^n) + n\delta\\
    &=I(M;Y^n|S^n,M_{f_1}) + n\delta\\
    &=\sum_{i=1}^n I(M;Y_i|Y^{i-1},S_i,S^{n\backslash i},M_{f_1})+ n\delta\\
    &\leq \sum_{i=1}^n I(S^{n\backslash i},M;Y_i|S_i,Y^{i-1},M_{f_1})+ n\delta\\
    &\overset{(a)}{\leq} \sum_{i=1}^{n} I(X_i;Y_i|S_i,Y^{i-1},M_{f_1})+ n\delta\\
    &\leq \sum_{i=1}^{n} I(X_i;Y_i|S_i,V_{11,i})+ n\delta\\
    &=I(X;Y|S,V_{11}),
\end{align*}
where $(a)$ follows by the Markov chain $(S^{n\backslash i},M,M_{f_1},Y^{i-1})-(S_i,X_i)-Y_i$.

By the definitions of the auxiliary random variables $V_{11}=(M_{f_1},Q)$ and $V_{12}=(S^{Q-1},Y_{Q+1}^n,T_Q,V_{11,Q},Q)$ and the distortion constraint \eqref{inq: converse distortion constraint}, we define the reconstruction function for each component of the reconstructed sequence by a new function $\widetilde{h}$ as
\begin{align}
    \hat{S}_i &= h_i(f_1(S^n),X^n,Y^n)\\
    &=h_i(f_1(S^n),X^{n\backslash i},Y^{i-1},Y^n_{i+1},X_i,Y_i)\\
    &=h_i((f_1(S^n),Y^{i-1}),X^{n\backslash i},Y^n_{i+1},X_i,Y_i)\\
    \label{def: converse single letter estimator}&:=\widetilde{h}((f_1(S^n),Y^{i-1},X^{n\backslash i},Y^n_{i+1},i),X_i,Y_i)\\
    &=\widetilde{h}(V_{12,i},T_i,X_i,Y_i)
\end{align}
It follows that
\begin{align}
    D &\geq \mathbb{E}\left[ d(S^n,\hat{S}^n) \right]\\
    &=\frac{1}{n}\sum_{i=1}^n \mathbb{E}\left[ d(S_i,\hat{S}_i) \right] \\
    &=\mathbb{E}\left[ \mathbb{E}\left[ d(S_Q,\hat{S}_Q) \right]| Q \right]\\
    &=\mathbb{E}\left[ d(S_Q,\hat{S}_Q) \right]\\
    &=\mathbb{E}\left[ d(S,\widetilde{h}(V_{12},T,X,Y)) \right],
\end{align}
where the last equality follows by the definition of $\widetilde{h}$ from \eqref{def: converse single letter estimator}.

\subsection{Converse of Corollary \ref{coro: causal state encoder}}\label{app: converse of causal state encoder}

The bound on $R$ follows the standard argument. For $R_{f_1}$, we have
\begin{align}
    R_{f_1} &\geq H(M_{f_1}|X^n,Y^n)\\
    &\geq I(M_{f_1};S^n|X^n,Y^n)\\
    &=\sum_{i=1}^n I(M_{f_1};S_i|X^n,Y^n,S^{i-1})\\
    &=\sum_{i=1}^n  H(S_i|X^n,Y^n,S^{i-1}) - H(S_i|X^n,Y^n,S^{i-1},M_{f_1})\\
    &=\sum_{i=1}^n H(S_i|X_i,Y_i,X^{n\backslash i}, Y^{n \backslash i},S^{i-1}) - H(S_i|X_i,Y_i,X^{n\backslash i}, Y^{n \backslash i},S^{i-1},M_{f_1})\\
    &\overset{(a)}{=}\sum_{i=1}^n H(S_i|X_i,Y_i,T_i) - H(S_i|X_i,Y_i,T_i,V_i),
\end{align}
where $(a)$ follows by setting $T_i=(X^{n\backslash i}, Y^{n \backslash i},S^{i-1})$ and $V_i=(T_i,M_{f_1})$.
To bound $R_{f_2}$, it follows that
\begin{align}
    R_{f_2} &\geq H(M_{f_2})\\
    &\geq H(M_{f_2}|M_{f_1},X^n)+H(Y^n|M_{f_1},M_{f_2},X^n) - H(Y^n|M_{f_1},M_{f_2},X^n)\\
    &\geq H(Y^n|M_{f_1},X^n) - n\epsilon\\
    &\geq \sum_{i=1}^n H(Y_i|M_{f_1},X_i,X^{n\backslash i},Y^{n \backslash i},S^{i-1})\\
    &=\sum_{i=1}^n H(Y_i|X_i,T_i,V_i).
\end{align}
For the sum rate, we have
\begin{align}
    R_{f_1} + R_{f_2} &\geq H(M_{f_1},M_{f_2})\\
    &\geq I(M_{f_1},M_{f_2};S^n,Y^n|X^n)\\
    &=I(M_{f_1},M_{f_2};Y^n|X^n) + I(M_{f_1},M_{f_2};S^n|X^n,Y^n)\\
    &=H(Y^n|X^n) - H(Y^n|X^n,M_{f_1},M_{f_2}) + H(S^n|X^n,Y^n) - H(S^n|X^n,Y^n,M_{f_1},M_{f_2})\\
    &=\sum_{i=1}H(Y_i|X_i) + \sum_{i=1}^n H(S_i|Y_i,X_i) - \sum_{i=1}^n H(S_i|X^n,Y^n,M_{f_1},M_{f_2},S^{i-1})- H(Y^n|X^n,M_{f_1},M_{f_2})\\
    &\overset{(a)}{=}\sum_{i=1}^n H(Y_i|X_i) + H(S_i|Y_i,X_i) - H(S_i|X^n,Y^n,M_{f_1},M_{f_2},S^{i-1}) - n\epsilon\\
    &\geq \sum_{i=1}^n H(Y_i|X_i) + H(S_i|Y_i,X_i) - H(S_i|X^n,Y^n,M_{f_1},S^{i-1}) - n\epsilon\\
    &=\sum_{i=1}^n H(Y_i|X_i) + H(S_i|Y_i,X_i) - H(S_i|X^{n \backslash i},Y^{n \backslash i},X_i,Y_i,M_{f_1},S^{i-1}) - n\epsilon\\
    &\overset{(b)}{=}\sum_{i=1}^n H(Y_i|X_i) + H(S_i|Y_i,X_i) - H(S_i|X_i,Y_i,V_i,T_i) - n\epsilon
\end{align}
where $(a)$ follows by the fact that $H(Y^n|X^n,M_{f_1},M_{f_2})\leq n\epsilon$ and Fano's inequality, $(b)$ follows by the definitions of $V_i$ and $T_i$.
The existence of the deterministic function $h$ can be proved by the definition of $V_i,T_i,i=1,2,...,n.$
Applying the standard time-sharing argument completes the converse proof.
\section{achievability of theorem \ref{the: common component inner and outer bounds}}\label{sec: inner bound proof of common components}
Fix a joint distribution $P_{S}P_{V_{11}|S}P_{T|V_{11}K}P_{V_{12}|TV_{11}S}P_{X|V_{11}}P_{Y|XS}$.

\emph{Codebook Generation at the state helper:} 
\begin{itemize}
    \item Upon observing the channel state $S^n$, the state helper uses function $c_1$ to generate the common component sequence $K^n=c_1(S^n)$.
    \item The state helper generates $N_{11}=\exp\{n(I(S;V_{11})+\epsilon)\}$ independent codewords $\{v^n_{11}\}$ for some $\epsilon>0$ indexed by $1\leq n_{11}\leq N_{11}$, each according to the distribution $P_{V_{11}}$.
    \item For each $v^n_{11}$, generate $N_{0}=\exp \{n(I(T;K|V_{11})+\epsilon)\}$ independent codewords $\{t^n|n_{11}\}$ according to $P_{T|V_{11}}$, indexed by $1\leq n_{0} \leq N_{0}$. 
    \item For each $v^n_{11}$ and $t^n$, generate $\widetilde{N}_{12}=\exp\{n(\widetilde{R}_{f_{12}}+\epsilon)\}$ independent codewords $\{v_{12}^n|n_{11},n_{0}\}$, each according to
    \begin{align}
        P_{V_{12}|V_{11}T}^n(v_{12}^n|v_{11}^n(n_{11}),t^n(n_{0}))=\prod_{i=1}^n P_{V_{12}|V_{11}}(v_{12,i}|v_{11,i}(n_{11}),t_i(n_{0})).
    \end{align}
    Partition each codebook $\{v_{12}^n|n_{11},n_{0}\}$ into $N_{12} = \exp\{n(R_{f_{12}}+\epsilon)\}$ bins $\{v_{12}^n|n_{11},n_{0},n_{12}\}$ indexed by $(n_{11},n_{0},n_{12}), 1\leq n_{12} \leq N_{12}$. Now, each codeword $v^n_2\in\{v_{12}^n|n_{11},n_{0},n_{12}\}$ can be uniquely indexed by $(n_{0},n_{11},n_{12},l)$, where $l$ is the index of $v^n_{12}$ within the bin. Denote the bin of a given $v_{12}^n$ and $(n_{11},n_{0})$ by $b(v^n_{12}|n_{11},n_{0})$.
\end{itemize}

\emph{Codebook Generation at the sender: } For each $v^n_{11}$, the sender generates $\widetilde{M}=\exp\{n(I(X;Y|S,V_{11})+\epsilon)\}$ independent codewords $\{x^n|n_{11}\}$, each according to $P_{X|V_{11}}(\cdot|v^n_{11}(n_{11}))$.

\emph{Codebook Generation at the feedback helper: } The feedback helper randomly assign an index $n_2\in[1:\exp\{n(R_{f_2}+\epsilon)\}]$ to each $y^n\in\mathcal{Y}^n$. Sequences with the same index form a bin. Denote the bin number of a given $y^n$ by $b(y^n)$.

\emph{Encoding at the state helper: } The indices sent by the state helper consist of three parts, which are denoted by $(\phi_{f_{11}}(s^n),\phi_{0}(s^n),\phi_{f_{12}}(s^n))$. 
\begin{itemize}
    \item At the beginning of each transmission block, the state helper observes the channel state $s^n$. It finds a codeword $v^n_{11}$ in $\{v^n_{11}\}$ with index $n_{11}$ such that $(s^n,v^n_{11})\in \mathcal{T}^n_{\delta}$. If there are multiple such codewords, the state helper selects the first one and then uses $n_{11}$ as the first index $\phi_{f_{11}}(s^n)$. If such a codeword does not exist, it sets $\phi_{f_{11}}(s^n)$ equal the default number 1 and declares an error.
    \item After finding $v^n_{11}(n_{11})$, the state helper computes a sequence $k^n=c_1(s^n)$ and then finds a $t^n\in\{t^n|n_{11}\}$ such that
    \begin{align}
        (v^n_{11},t^n,k^n)\in \mathcal{T}^n_{\delta}
    \end{align}
    If there are multiple such $t^n$, then the state helper chooses the first one. If there is no such sequence, the state helper sets $\phi_{0}(s^n)$ equal to the default number 1 and declares an error.
    It uses the index of $t^n$ in $\{t^n|n_{11}\}$ as the second index $\phi_{0}(s^n)$. 

    \item After finding $(v^n_{11}(n_{11}),t^n(n_{11},n_0))$, the state helper then finds a codeword $v^n_{12}\in\{v^n_{12}|n_{11},n_0\}$ such that
    \begin{align}
        (t^n,v^n_{11},v^n_{12},s^n)\in\mathcal{T}^n_{\delta}.
    \end{align}
    If there are multiple such codewords, then the state helper selects the first one. If such a codeword does not exist, the default number is selected, and the state helper declares an error. The third index $\phi_{f_{12}}(s^n)$ sent by the state helper is the bin $b(v^n_{12}|n_{11},n_{0})$ of the selected codeword $v^n_{12}$. For simplicity, we also denote the selected codewords given $s^n$ by $v^n_{11}(s^n),t^n(s^n)$ and $v^n_{12}(s^n)$.
\end{itemize}

\emph{Encoding at Sender: } The sender receives three indices $(\phi_{f_{11}}(s^n),\phi_{f_{0}}(s^n),\phi_{f_{12}}(s^n))$ from the state helper at the beginning of the block. It finds $v^n_{11}$ according to the first index $\phi_{f_{11}}(s^n)$. To transmit message $m$, it use codeword $x^n(n_{11},m)$ from the codebook $\{x^n|n_{11}\}$. 

\emph{Encoding at the feedback helper:} At the end of the transmission block, the feedback helper receives all the feedback symbols $y^n$. It set $\phi_{f_2}(y^n)=b(y^n)$. 

\emph{Decoding at the sender: } Although the sender receives three indices from the state helper at the beginning of the transmission block, it can only use the first two $(\phi_{f_{11}}(s^n),\phi_0(s^n))$ to locate the codeword $(\hat{v}^n_{11},t^n)$. It is not able to decode $\phi_{f_{12}}(s^n)$ due to the lack of side information until it selects the codeword and receives $\phi_{f_2}(y^n)$ at the end of the current transmission block.

Then, the decoder looks for a $\hat{y}^n\in \{y^n|\phi_{f_2}(y^n)\}$ such that
\begin{align}
    (\hat{t}^n,\hat{v}^n_{11},x^n,\hat{y}^n)\in\mathcal{T}^n_{\delta}
\end{align}

It finally tries to find a unique $\hat{v}^n_{12}$ in the bin $\{v^n_{12}|\phi_{f_{11}}(s^n),\phi_0(s^n),\phi_{f_{12}}(s^n)\}$ such that
\begin{align}
    (\hat{t}^n,\hat{v}^n_{11},x^n,y^n,\hat{v}^n_{12})\in\mathcal{T}^n_{\delta}.
\end{align}

\emph{Decoding at the receiver: } The receiver observes the channel output $y^n$ and state sequence $s^n$. It looks for a unique $\hat{x}^n$ in the communication codebook such that $(v^n_{11},x^n,y^n,s^n)\in\mathcal{T}^n_{\delta}$. If such a codeword does not exist or there are multiple such codewords, the receiver declares an error. 

\emph{State Estimation: } The sender decodes codewords $(v^n_{11},t^n,v^n_{12},v^n_{2})$ by the indices $(\phi_{f_{11}}(s^n),\phi_{f_{0}}(s^n),\phi_{f_{12}}(s^n),\phi_{f_2}(y^n))$ sent by the state and feedback helpers. It reproduces the state sequences by $\hat{s}(x,v_{12},y).$

\textbf{Error Analysis: } In the following, we use $B_{0},B_{12},B_2$ as the bin indices of codewords $(T^n,V^n_{12},Y^n)$ and $L_{10},L_{12},L_2$ as the indices of $(T^n,V^n_{12},Y^n)$ within the bins. We define error events as follows.

\emph{Encoding error at the state helper:}
\begin{align*}
    &\mathcal{E}_{11} =\left\{ \bigcap_{n_{11}=1}^{N_{11}}(V^n_{11}(n_{11}),S^n)\notin \mathcal{T}^n_{\delta}  \right\},\\
    &\mathcal{E}_{12} = \left\{ \bigcap_{n_{10}=1}^{N_{10}}(V^n_{11}(S^n),T^n(n_{10}),K^n)\notin \mathcal{T}^n_{\delta}   \right\}\\
    &\mathcal{E}_{13} = \left\{ \bigcap_{\widetilde{n}_{12}=1}^{\widetilde{N}_{12}}(V^n_{11}(S^n),T^n(S^n),V^n_{12}(\widetilde{n}_{12}),S^n)\notin \mathcal{T}^n_{\delta}  \right\}.
\end{align*}
To bound the error event $\mathcal{E}_{12}$, with the setting of $N_{11},\widetilde{N}_0$, we invoke the generalized conditional Markov lemma\cite[Lemma 9]{wagner2011distributed}. The error probability of $\mathcal{E}_{12}$ vanishes with $n\to\infty$ if we have
\begin{align}
    \widetilde{R}_{f_{12}} > I(V_{12};S|V_{11},T). 
\end{align}
The bound of $\mathcal{E}_{11}$ and $\mathcal{E}_{13}$ follow similarly to the analysis of the coding scheme for Theorem \ref{the: inner bound}.

\emph{Decoding error at the sender:}
\begin{align}
    &\mathcal{E}_2 = \{(T^n(S^n),V^n_{11}(S^n),V^n_{12}(S^n),X^n(M),S^n,Y^n) \notin \mathcal{T}^n_{\delta}\},\\
    &\mathcal{E}_3 = \{(T^n(S^n),V^n_{11}(S^n),X^n,V^n_{12}(B_{12},L_{12}'), Y^n)\in\mathcal{T}^n_{\delta}\;\text{for some $L_{12}'\neq L_{12}$}\},\\
    &\mathcal{E}_4 = \{(T^n(S^n),V^n_{11}(S^n),X^n,Y^n(B_2,L_2'))\in\mathcal{T}^n_{\delta}\;\text{for some $L_{2}'\neq L_{2}$}\}.
\end{align}
By the conditional typical lemma, we have
\begin{align}
    Pr\{\mathcal{E}_2\} \to 0.
\end{align}
The probability of $\mathcal{E}_3$ can be proved similar to the Wyner-Ziv problem using the packing lemma. The error probability vanishes with $n\to\infty$ if we have
\begin{align}
    R_{f_{12}} \geq I(V_{12};S|T,V_{11}) - I(V_{12};X,Y|T,V_{11}) = I(V_{12};S|T,X,Y,V_{11}).
\end{align}
For $\mathcal{E}_4$, it follows that
\begin{align}
    Pr\{\mathcal{E}_4\} &= Pr\{\mathcal{E}_4|b(Y^n)=B_2\}\\
    &=\sum_{t^n,v^n_{11},x^n,y^n}P(t^n,v^n_{11},x^n,y^n|b(Y^n)=B_2)\\
    &\quad\quad\quad\quad P({y^n}'\;\text{s.t. $b({y^n}')=B_2,{y^n}'\neq y^n,(t^n,v^n_{11},x^n,{y^n}')\in\mathcal{T}^n_{\delta}$}|t^n,v^n_{11},x^n,y^n,b(y^n)=b_2)\\
    &\leq \sum_{\substack{{y^n}'\neq y^n,\\(t^n,v^n_{11},x^n,{y^n}')\in\mathcal{T}^n_{\delta}}} P(b({y^n}'=B_2))\\
    &\leq 2^{n(H(Y|T,X,V_{11})+\epsilon)}2^{-nR_{f_{2}}}.
\end{align}
It follows that the probability of $\mathcal{E}_5$ can be made arbitrarily small if we have
\begin{align}
    R_{f_2} \geq H(Y|T,X,V_{11}).
\end{align}
\section{Conclusion}
This paper studies integrated sensing and communication when the sender and estimator are assisted by some rate-limited helpers. The considered problem aims to model real-world communication and sensing in a network such that each participant dynamically acts as a sender, receiver, or sensor. Inner bound of the capacity-compression-distortion region and some capacity-achieving special cases are provided. 

The ISAC models analyzed in the paper represent simplifications of future practical ISAC systems. Nevertheless, the proposed coding scheme provides insights into the design of real-world systems. The existence of the helpers increases the degree of freedom of the tradeoff by introducing compression into the problem. What's more, the compression itself incorporates a tradeoff depending on how to use the information that the helper has. In addition to the improvement of the sensing quality, we show by numerical examples that the helpers facilitate the communication even when the sender cannot decode all of the information they send in the encoding phase.

\bibliographystyle{ieeetr} 
\bibliography{ref}
\end{document}